# Calibrated Photoacoustic Spectrometer Based on a Conventional Imaging System for In Vitro Characterization of Contrast Agents

**Théotim Lucas [1,2,†], Mitradeep Sarkar [3,†], Yoann Atlas [1], Clément Linger [1,4], Gilles Renault [5], Florence Gazeau [2] and Jérôme Gateau [1,*]**

[1] Laboratoire d'Imagerie Biomédicale, Sorbonne Université, CNRS, INSERM, LIB, 75006 Paris, France
[2] Matière et Systèmes Complexes, Université Paris Cité, CNRS, MSC, 75006 Paris, France
[3] Paris Cardiovascular Research Center, Université Paris Cité, INSERM, PARCC, 75015 Paris, France
[4] Institut Galien Paris-Saclay, Université Paris-Saclay, CNRS, IGPS, 91400 Orsay, France
[5] Institut Cochin, Université Paris Cité, INSERM, CNRS, 75014 Paris, France
* Correspondence: jerome.gateau@sorbonne-universite.fr; Tel.: +33-144272265
† These authors contributed equally to this work.

**Abstract:** Photoacoustic (PA) imaging systems are spreading in the biomedical community, and the development of new PA contrast agents is an active area of research. However, PA contrast agents are usually characterized with spectrophotometry or uncalibrated PA imaging systems, leading to partial assessment of their PA efficiency. To enable quantitative PA spectroscopy of contrast agents *in vitro* with conventional PA imaging systems, we have developed an adapted calibration method. Contrast agents in solution are injected in a dedicated non-scattering tube phantom imaged at different optical wavelengths. The calibration method uses a reference solution of cupric sulfate to simultaneously correct for the spectral energy distribution of excitation light at the tube location and perform a conversion of the tube amplitude in the image from arbitrary to spectroscopic units. The method does not require any precise alignment and provides quantitative PA spectra, even with non-uniform illumination and ultrasound sensitivity. It was implemented on a conventional imaging setup based on a tunable laser operating between 680 nm and 980 nm and a 5 MHz clinical ultrasound array. We demonstrated robust calibrated PA spectroscopy with sample volumes as low as 15 μL of known chromophores and commonly used contrast agents. The validated method will be an essential and accessible tool for the development of new and efficient PA contrast agents by improving their quantitative characterization.

## 1. Introduction

Photoacoustic imaging (PAI) is an emerging multi-wave biomedical imaging modality able to reveal molecular information at centimeter depths in biological tissues and with sub-millimeter resolutiaon [1]. PAI is based on the photoacoustic (PA) effect: optically absorbing materials emit ultrasound waves when excited with transient illumination. The ultrasound waves are generated by thermoelastic expansion, and their amplitude is proportional to the absorbed optical energy at the excitation wavelength. Therefore, successive acquisitions of PA images at different optical wavelengths may allow spectral discrimination and quantification of the various absorbers in the imaged region [2].

To enhance this hybrid imaging modality beyond the information provided by endogenous absorbers such as hemoglobin, the injection of absorbing exogenous contrast agents is often required [3]. Recently, the material science community has shown a growing interest in the development of novel PA contrast agents [4,5], resulting in a strong need for techniques able to characterize them in terms of effective PA spectra and efficiency to generate ultrasound. Spectrophotometry (SPP), based on the transmission of light by a sample, usually measures the optical attenuation: the sum of losses due to the absorption and the scattering of light. However, the latter does not contribute to PA signal generation. Moreover, SPP does not account for the photophysical and thermoelastic processes that occur during optical absorption and the subsequent ultrasound pressure generation.

For many of the developed PA contrast agents, the *in vivo* detectability with PAI is demonstrated using commercial PAI systems or conventional prototypes. Therefore, PAI systems are available to many research groups and could be advantageously used to perform a more quantitative *in vitro* PA characterization of the agents than with SPP. Commercial PAI systems [6,7] have already been proposed to measure the PA spectral response of contrast agents. However, no calibration was performed for such systems, leading to PA spectral assessments in arbitrary units. On the other hand, dedicated calibrated PA spectrometers have also been developed. However, they do not use a PAI system and therefore require a specific instrument. Beard et al. [8–11] developed a PA spectrometer able to measure the absolute optical absorption coefficient by fitting an analytic expression to the photoacoustically generated ultrasound signal. Furthermore, photoacoustic specific coefficients could be calculated with this system: the photothermal conversion efficiency $E_{pt}$, which represents the conversion efficiency of the absorbed optical energy to heat, and the Grüneisen coefficient $\Gamma$ (relative to water), which describes the conversion of the heat energy to the initial pressure rise resulting in the ultrasound waves. However, this PA spectrometer requires large sample volumes (mL) and a specific ultrasound detector with a very broadband and flat frequency response to correctly resolve the ultrasound waveform [10]. Other PA spectrometers based on dedicated single-element detectors place fewer constraints related to the ultrasound frequency response of the detector. They evaluate the optical absorption coefficient using calibration with a known reference solution [12,13]. For these PA spectrometers, small sample volumes (3 µL [13] and 200 µL [12]) are placed in optically transparent cells, and SPP is performed on the same sample to concurrently measure the optical attenuation.

We have developed and we present herein a calibration method using a reference solution to transform a conventional multispectral PAI system into a calibrated PA spectrometer for the *in vitro* characterization of PA contrast agents. We implemented the method in a conventional configuration for PAI [14,15]: a clinical linear ultrasound detector array with light delivered from the side. For PA contrast agent characterization, small sample volumes (15 µL) were injected in tubes whose diameter was chosen such that the ultrasound emission matches the frequency bandwidth of the detector. We demonstrate that our simple experimental setup enables robust calibrated spectroscopic measurements of photoacoustic contrast agents.

## 2. Materials and Methods

### 2.1. Experimental Setup and Data Acquisition

The experimental setup is presented in Figure 1a. It is comprised of a sample compartment and a conventional multispectral PAI system.

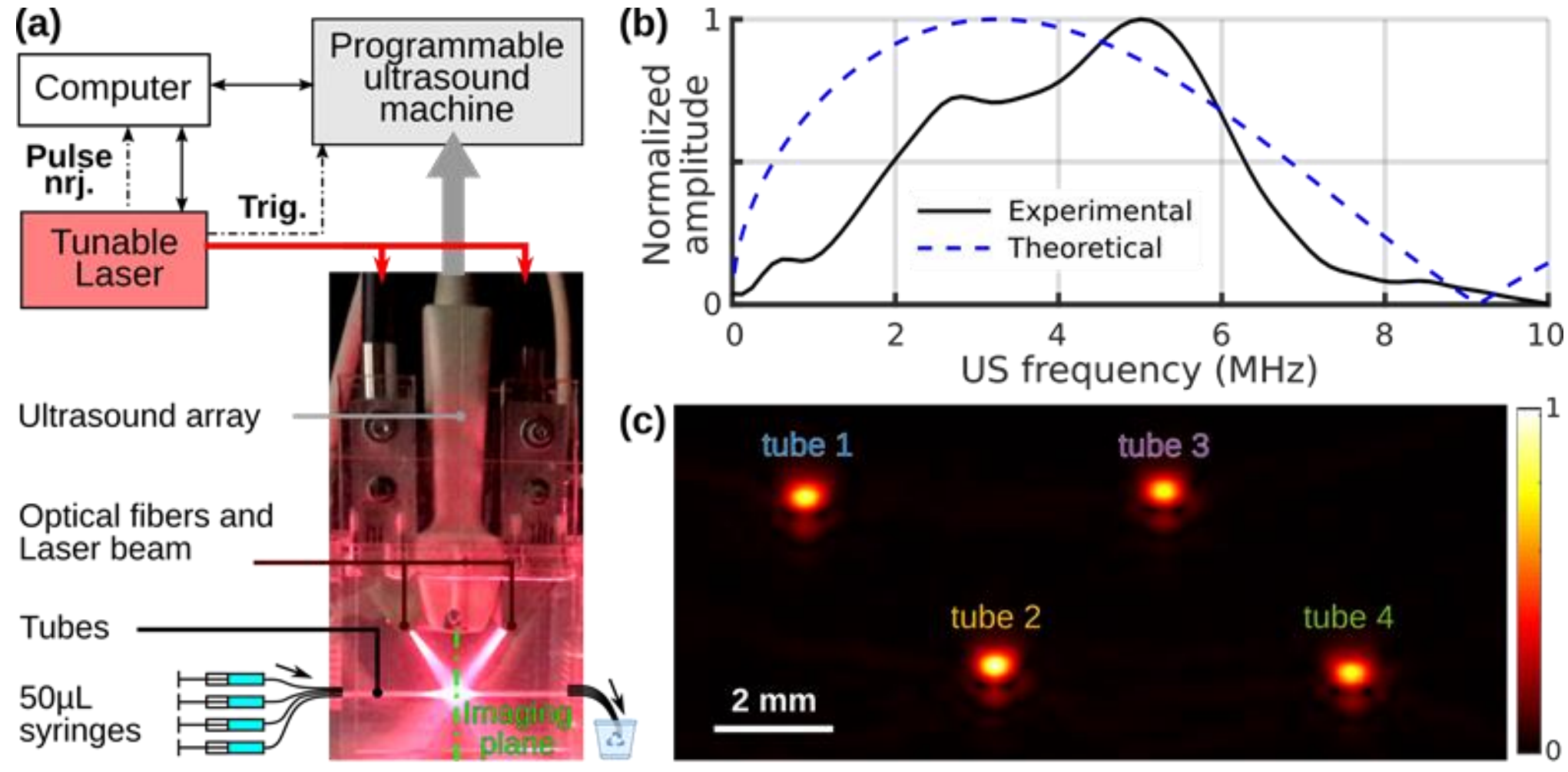

**Figure 1.** Experimental setup: (**a**) Annotated picture of the PAI system and schematic drawing of the experimental setup. The imaging plane of the array is perpendicular to the picture plane and perpendicular to the tubes. The laser beam was especially made visible for the picture with dust particles added in the water bath. (**b**) Experimental and theoretical frequency spectra of the PA-generated ultrasound signal. The experimental ultrasound spectrum was acquired with a tube filled with the calibration solution at 710 nm. The theoretical spectrum corresponds to Equation (A1). The spectra are normalized to their maximum values. (**c**) Image of 4 tubes filled with the calibration solution.

#### 2.1.1. Sample Compartment

The sample compartment consists of 50 cm long polytetrafluoroethylene (PTFE) tubes (inner diameter: 0.2 mm, wall thickness: 0.1 mm, Bola, Germany). PTFE (Teflon) is hydrophobic and chemically inert; therefore, PTFE tubes are well adapted to contain aqueous solution of PA contrast agents. Moreover, PTFE was shown to have a weak optical absorption in the near infrared (NIR) [16] to avoid strong background PA signals. The tubes were threaded through holes of two parallel perforated plates located 8 cm apart. The tubes were arranged to be parallel one to the other. Up to four tubes were positioned in the sample holder to perform simultaneous data acquisition (Figure 1c). The tubes were positioned in a staggered arrangement instead of an in-line arrangement to avoid both optical shadowing and interferences of the emitted ultrasound waves at the detector surface. To ensure a good spatio-temporal separation of the ultrasound signals from each tube and thereby independent measurements, a distance of at least 4 mm between the tubes was arbitrary chosen here. This distance may be reduced for a higher number of tubes. The tubes were immersed in a water tank (ultrapure water, resistivity 18 MΩ·cm, Purelab Option Q, ELGA LabWater) at room temperature to ensure acoustic coupling between the samples and the ultrasound detector of the PAI system. The two ends of each tube were kept out of the water tank to inject and collect the measured samples, respectively. The inner volume of each tube was 15 µL, and the tubes were filled using a 33-gauge needle and a 50 µL gas-tight syringe (Hamilton). A thermometer (HI98509, Hanna instruments, Lingolsheim, France) was used to monitor the temperature of the water bath with a precision of ±0.2 °C. The temperature of the bath had a maximum variation range of 1 °C for each series of measurement.

#### 2.1.2. The Conventional PAI System

A tunable (680–980 nm) optical parametric oscillator laser (SpitLight 600 OPO, Innolas Laser GmbH, Krailling, Germany) delivering < 8 ns pulses with a pulse repetition frequency of 20 Hz was used to generate the optical excitation. The tuning range corresponds to the first optical window in biological tissue and to the near-infrared wavelength range typically used in PA tomography. Ultrasound was detected with a 128-element clinical linear array (L7-4, 5 MHz center frequency, bandwidth 4–7 MHz, ATL) driven by a programmable ultrasound machine used in receive-only mode (Vantage, Verasonics, WA, USA). A bifurcated fiber bundle (CeramOptec GmbH, Bonn, Germany) guided the light toward the fixed elevational focus of the ultrasound array (located at 25 mm from the surface of the detector) and delivered light over the entire length of the array. The tubes containing the samples were placed perpendicularly to the imaging plane of the ultrasound detector and near the elevational focus for higher sensitivity. Therefore, the intersection of the imaging plane with each tube was a disk. The illuminated length (perpendicular to the imaging plane) was around 1.5 cm. Figure 1b shows that the ultrasound frequency bandwidth of the detector (4–7 MHz) is fully included in the main peak of the ultrasound spectrum generated by one tube (Appendix A). Then, the detection sensitivity is high for this sample container.

The mean fluence at the elevational focus was estimated to be around 4 mJ.cm$^{-2}$ at 730 nm (wavelength at which the laser has the maximum pulse energy). This estimation was performed in air by dividing the laser energy per pulse (Pyroelectric Energy sensor ES245C, Thorlabs, Newton, NJ, USA) by the area covered by the excitation light in the elevational focal plane (at 25 mm of the detector surface). The fluence was lowered for some samples by reducing the laser energy before injection in the fiber bundle thanks to polarizing optics. Because of broken fibers, the finite length of the outputs of the fiber bundle, and the staggered arrangement of the tubes, the laser fluence was not uniform over the tubes. However, neither the precise knowledge of the laser fluence nor the uniformity of the illumination is required for our method since a calibration procedure is carried out per tube and for each series of PA spectrum measurements.

#### 2.1.3. Measurement Process

For the PA acquisition, each laser pulse triggered (1) an ultrasound acquisition in parallel on all the 128 elements of the detector array and (2) a recording of the pulse energy using a pyrometer incorporated in the laser. The incorporated pyrometer was not calibrated, but it was verified, using an external calibrated pyroelectric energy meter (PE50BF-DIFH-C, Ophir Photonics), that the delivered electric signal was proportional to the pulse energy for each laser pulse at a given wavelength. For a spectroscopic acquisition, measurements were performed successively at different optical wavelengths ($\lambda$) over the entire tunable spectral range of the laser and at an acquisition rate of 20 Hz. The per-pulse tunability of the laser was used, and the wavelength sequence was programmed with the laser software interface. The acquisition sequence consisted in recording the ultrasound signals and the corresponding pyrometer values for 15 successive sweeps of 30 wavelengths between 680 nm and 970 nm with a step of 10 nm, for a total of 30 × 15 = 450 laser pulses. This swept sequence avoids consecutive excitations at a given wavelength that could induce photodegradation. Any potential changes in the PA spectra of the sample during the acquisition sequence can be detected as the entire spectral range is covered 15 times consecutively. For all samples reported in this paper, the spectra were found to be stable during the experimental sequences. Therefore, iterations at a given wavelength were averaged to increase the signal-to-noise ratio. Before averaging, ultrasound signal amplitudes were simply divided by the corresponding pyrometer value to correct for the pulse-to-pulse energy fluctuations of the laser.

For measurements with a spectrophotometer (SPP), a baseline correction is performed using a "blank" measurement obtained by filling the sample compartment with the solvent. In a similar manner, for our PA spectrometer, a blank dataset was acquired with the tubes filled with ultrapure water (or the solvent when available). For baseline correction and suppression of the background signal of the tubes, the blank dataset was coherently subtracted from the averaged signals of the tube filled with the sample (subtraction of the radio-frequency signals). Thereby, the effective signals from the contrast-agent sample were isolated. During a series of measurements, blank datasets were recorded on a regular basis (between two different samples) to ensure an accurate correction and, at the same time, to verify that the tube was not polluted by a sample (comparison with a previous blank dataset to check for sample-induced persistent absorption). Following the baseline correction, the Hilbert transform of the corrected signals was computed to obtain quadrature signals. The in-phase signals and quadrature signals were beamformed independently using a simple delay-and-sum image reconstruction algorithm to yield two images. The speed of sound in the water bath was estimated using the measured temperature [17]. Then, an envelope-detected image was computed from the root-mean square of the two images for each pixel. The envelope-detected image of a sample injected in four tubes is presented in Figure 1c. Each tube appeared as a Gaussian spot, and its amplitude $A^{PA}$ $(\lambda)$ was determined using a 2D Gaussian fit. $A^{PA}$ $(\lambda)$ depends on the tube, the sample, and the optical wavelength $\lambda$.

### *2.2. Calibrated Measurements with the PA Spectrometer*

The calibration of the PA spectrometer aims at retrieving a PA spectrum $\theta^{PA}(\lambda)$ in the spectroscopic units of the absorption coefficient $\mu_a(\lambda)$, typically $cm^{-1}$, from $A^{PA}$ $(\lambda)$, while avoiding any tedious calibration of the ultrasound detector, determination of the light fluence distribution, or any precise alignment. We base our calibration process on a calibration per sample container and per acquisition series using a calibration solution injected in the tube prior to the samples in the series of measurements.

#### 2.2.1. The Photoacoustic Coefficient of a Sample $\theta^{PA}(\lambda)$

For the laser pulse width and the dimension of the tube used here, the thermal and stress confinement regimes are satisfied [18]. Therefore, the thermal expansion of the sample inside the tube is expected to cause a pressure rise $p_0$ proportional to the absorption coefficient $\mu_a(\lambda)$ of the sample:

$$p_0(\lambda) = \Phi(\lambda) \cdot \Gamma_{water} \cdot \eta_{sample}(\lambda) \cdot \mu_a(\lambda), \tag{1}$$

where $\Phi(\lambda)$ is the local light fluence at the tube location for the wavelength $\lambda$, and $\Gamma_{water}$ is the Grüneisen coefficient of water. Water is the main constituent of the aqueous solutions used here. $\eta_{sample}$ is the dimensionless photoacoustic generation efficiency (PGE) of the sample. It corresponds to the efficiency of the PA pressure generation compared to a sample for which the absorbed energy is fully converted into pressure in a medium with the Grüneisen coefficient of the water. According to the conventional photoacoustic theory [19], $\eta_{sample}$ can be expressed as:

$$\eta_{sample}(\lambda) = E_{pt,sample}(\lambda) \cdot \Gamma_{sample}/\Gamma_{water}, \tag{2}$$

where $E_{pt,sample}$ and $\Gamma_{sample}$ are the photothermal conversion efficiency and the Grüneisen coefficient of the sample solution, respectively. The photothermal conversion efficiency $E_{pt,sample}$ is the ratio of the energy effectively converted into a thermal increase of the solution (and subsequently to production of ultrasound waves) to the total absorbed optical energy. $E_{pt,sample}$ may be inferior to 1, due to various competitive pathways [9]. For molecular absorbers, fluorescence and other energy-transfer mechanisms [12] can attenuate the conversion efficiency. A realistic value for the Grüneisen coefficient of water taken from the literature is $\Gamma_{water} = 0.12$ at 22 °C [20], but the presence of solute and salts may increase the Grüneisen coefficient $\Gamma_{sample}$ of the aqueous solution.

In our experimental setup, $\Phi(\lambda)$ and $\Gamma_{water}$ are independent of the sample placed in the tube. Therefore, the sample-dependent factors of Equation (1) can be isolated in a quantity $\theta^{PA}(\lambda)$, named here the PA coefficient of the sample:

$$\theta^{PA}(\lambda) = \eta_{sample}(\lambda) \cdot \mu_a(\lambda), \tag{3}$$

The photoacoustic coefficient corresponds to the optical absorption coefficient of the sample, restricted to the absorption effectively transferred into heat to the surroundings of the absorbing agent and corrected for the potentially modified Grüneisen coefficient compared to the main solvent: water.

For non-scattering solutions, the absorption coefficient $\mu_a(\lambda)$ is equal to the attenuation coefficient $\mu^{SPP}(\lambda)$ and can be directly measured with spectrophotometry (SPP) in transmission mode. Using the measured absorbance $Abs^{SPP}$ $(\lambda)$ of a solution, $\mu^{SPP}(\lambda)$ was obtained with the following formula:

$$Abs^{SPP}(\lambda) = \frac{\mu^{SPP}(\lambda)}{ln(10)} \cdot L = \mu_{10}^{SPP}(\lambda) \cdot L, \tag{4}$$

where $L$ = 1 cm is the length of the SPP cuvette, and $ln$ is the natural logarithm. $\mu_{10}^{SPP}(\lambda)$is the decadic absorption coefficient.

### 2.2.2. Theoretical Relationship between $A^{PA}$ and $\theta^{PA}(\lambda)$

The amplitude of the ultrasound signal recorded with the ultrasound system is proportional to the pressure rise $p_0$ generated in the illuminated tube. Moreover, the "delay and sum" image reconstruction process used here is linear. Therefore, the amplitude $A^{PA}$ $(\lambda)$ computed from the image is proportional to $p_0$. Furthermore, the ultrasound signals were corrected for the pulse-to-pulse energy fluctuations of the laser. Therefore, the theoretical relationship between $A^{PA}$ and $\theta^{PA}(\lambda)$ can be expressed as:

$$A^{PA}(\lambda) = \nu_{US} \cdot \Phi_{avg}(\lambda) \cdot \nu_{Py}(\lambda) \cdot \Gamma_{water} \cdot \theta^{PA}(\lambda) \; = \delta_{tube}(\lambda) \cdot \theta^{PA}(\lambda), \tag{5}$$

where $\nu_{US}$ is a coefficient that accounts for the global conversion of $p_0$ in the arbitrary units of the beamformed image, $\Phi_{avg}(\lambda)$ is the pulse-average spectral energy distribution of the laser, and $\nu_{Py}(\lambda)$ is a coefficient that accounts for the pyrometer spectral sensitivity and the optical attenuation between the laser output and the tube. These factors can be gathered in a proportionality factor $\delta_{tube}(\lambda)$. $\delta_{tube}(\lambda)$ ensures the conversion between the arbitrary units of $A^{PA}$ and the spectroscopic units of $\mu_a(\lambda)$. $\delta_{tube}(\lambda)$ is independent of the sample but dependent of the optical wavelength and of the tube location in the image.

The calibration process consists in assessing $\delta_{tube}(\lambda)$ using a reference solution for which $A^{PA}{}_{calibration}$ $(\lambda)$, $\mu_a{}^{calibration}(\lambda)$, and $\eta_{calibration}(\lambda)$ have been predetermined:

$$\delta_{tube}(\lambda) = A_{calibration}^{PA}\,(\lambda) / \left(\eta_{calibration}(\lambda) \cdot \mu_a^{calibration}(\lambda)\right), \tag{6}$$

Each series of measurements started and ended with two acquisitions with the calibration solution per tube. $\tilde{A}^{PA}{}_{calibration}$ $(\lambda)$ was computed as the median value over the four acquisitions.

The choice of the calibration solution is presented in Section 2.3.2. We chose a calibration solution for which $\eta_{calibration}$ is independent of the optical wavelength. However, the determination of $\eta_{calibration}$ was not straightforward. Without prior knowledge of $\eta_{calibration}$, the following quantities were computed:

$$\Psi^{PA}(\lambda) = \frac{A^{PA}(\lambda)}{\tilde{A}_{calibration}^{PA}(\lambda)} \tag{7}$$

and

$$\xi^{PA}(\lambda) = \Psi^{PA}(\lambda) \cdot \mu_a^{calibration}(\lambda) = \frac{\theta^{PA}(\lambda)}{\eta_{calibration}} \tag{8}$$

For the sake of clarity, the notations of the different computed quantities are summarized in Figure 2 and listed in Table 1.

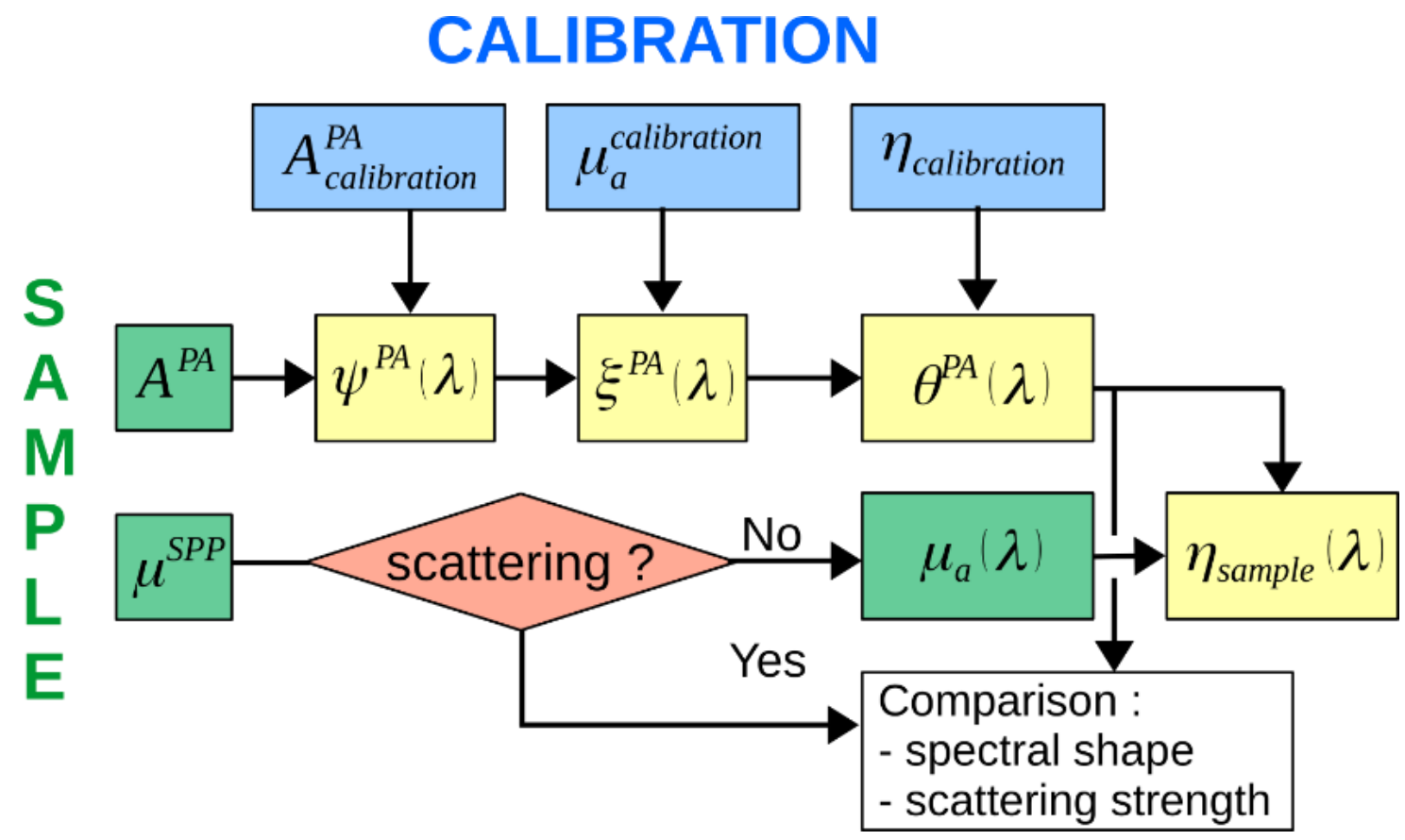

**Figure 2.** Diagram summarizing the computed quantities and their relationship to the measurements performed with the calibration solution and the sample solution. The symbols are listed in Table 1.

**Table 1.** List of symbols.

| Symbol | Name | Corresponding Equation or Section |
| --- | --- | --- |
| $\lambda$ | Optical wavelength | |
| $A^{PA}$ | Amplitude of one tube in the image | Section 2.1.3 |
| $\mu^{SPP}$ | Attenuation coefficient | Section 2.2.1 |
| $\mu_a$ | Absorption coefficient | |
| $\eta$ | Photoacoustic generation efficiency | Equation (2) |
| $\theta^{PA}$ | Photoacoustic coefficient | Equation (3) |
| $\Psi^{PA}$ | | Equation (7) |
| $\xi^{PA}$ | | Equation (8) |
| $\gamma$ | Fitting coefficient between $\Psi^{PA}$ and 1 | Section 3.1 |
| $\alpha$ | Fitting coefficient between $\xi^{PA}$ and $\mu_a$ | Equation (12) |

#### 2.2.3. Statistical Evaluation

Typically, 2 to 3 acquisitions in 4 tubes were performed to reach 8 to 12 measurements of the PA spectrum per sample. The median of the computed quantity was used as an estimate to avoid outliers due to possible injection errors. The median estimate was notated:

$$\tilde{X} = \text{median}(X) \tag{9}$$

For the evaluation of the measurement error, we used the median absolute deviation MAD, defined as:

$$\text{MAD}(X) = 1.4826 \cdot \text{median}\left( \left|X - \tilde{X}\right| \right) \tag{10}$$

The scale factor ensures that the value of MAD is comparable with the value of the standard deviation if the $X$ values are normally distributed.

When expected values are known, the median relative error (MRE) was computed. For example, $\xi^{PA}$ is expected to be equal to $\mu_a^{calibration}$ for an acquisition corresponding to the calibration solution. The MRE is defined as:

$$\text{MRE}(\xi^{PA}(\lambda), \mu_a(\lambda)) = \frac{1}{\mu_a(\lambda)} \cdot 1.4826 \cdot \text{median}(\,|\xi^{PA}(\lambda) - \mu_a(\lambda)|\,) \tag{11}$$

### 2.3. Preparation of the Absorbing Solutions

All solutions were prepared at room temperature. Two different groups of solutions were prepared: non-fluorescent and non-scattering molecular solutions and commonlyused PA contrast agents, which are either fluorescent or scattering.

#### 2.3.1. Non-Fluorescent and Non-Scattering Molecular Solutions

Three different water-soluble molecular absorbers with a photothermal conversion efficiency $E_P$ that can be assumed to be equal to 1 were selected for the characterization of the PA spectrometer: cupric sulfate pentahydrate ($CuSO_4{\cdot}5H_2O$, ACS reagent, ≥98.0%, Sigma-Aldrich, St. Louis, MO, USA), nickel sulfate hexahydrate ($NiSO_4{\cdot}6H_2O$, ACS reagent, ≥98.0%, Sigma-Aldrich, St. Louis, MO, USA), and nigrosin (Nigrosin, 198285, Sigma-Aldrich, St. Louis, MO, USA). These compounds absorb in the investigated optical range 680–970 nm. The assumption of $E_{pt} = 1$ means that all the absorbed optical energy is transferred to the solution into heat. To the best of our knowledge, there is no other de-excitation pathway for these compounds. For $CuSO_4{\cdot}5H_2O$ and $NiSO_4{\cdot}6H_2O$, the assumption of $E_{pt} =1$ has previously been reported for aqueous solutions of copper (II) chloride and nickel(II) chloride [9]. Nigrosin is non-florescent dye that was used to provide controlled optical absorption to tissue mimicking phantoms [21]. Nigrosin was excited away from its absorption peak, which is around 570 nm, and was found to be photostable.

Stock solutions were prepared to obtain a maximum absorption coefficient around $\mu_a$ = 7 cm$^{-1}$ in the wavelength range 680–970 nm. The stock solutions were prepared by adding crystals or powder in a 50.0 mL volumetric glass flask. The mass for each compound is given in Table 2. The flask was gradually filled with ultrapure water (resistivity 18 MΩ·cm, Purelab Option Q, ELGA LabWater) to dissolve the crystals/powder and obtain an accurate concentration. We observed a volume contraction when dissolving the sulfate crystals. Dilutions of the stock solution were performed to prepare percent solutions (*v*/*v* %): 100%, 80%, 60%, 40%, and 20%. A mix solution, named mix-$SO_4$, was also prepared, consisting of 30% (*v*/*v*) $CuSO_4{\cdot}5H_2O$ stock solution, 30% $NiSO_4{\cdot}6H_2O$ stock solution, and 40% water. Mixtures of the sulfate solutions with the nigrosin solution were found to form a precipitate and were not used.

**Table 2.** Parameters for the stock solutions.

| Compound | Mass of Solid for 50.0 mL of Solution | Molar Mass | Molar Concentration | Wavelength ($\lambda_{max}$) at the Absorption Maximum [1] | Molar Absorptivity at $\lambda_{max}$ | Relative Range [2] of $\mu_a$ [1] |
|---|---|---|---|---|---|---|
| $CuSO_4{\cdot}5H_2O$ | 3.12 g (crystals) | 249.69 g/mol | 250 mM | 810 nm | 12 $M^{-1}.cm^{-1}$ | 0.7 |
| $NiSO_4{\cdot}6H_2O$ | 18.00 g (crystals) | 262.85 g/mol | 1.37 M | 720 nm | 2.2 $M^{-1}.cm^{-1}$ | 1.9 |
| Nigrosin | 12 mg (powder) | 202.21 g/mol | 1.2 mM | 680 nm | ~$2.10^3$ $M^{-1}.cm^{-1}$ | 1.5 |

[1] The values are within the range of optical wavelengths 680–970 nm. [2] The relative range is ratio of the variation range (absolute difference between the maximum and minimum values) to the mean value.

The compounds have similar low molar mass (Table 2) and are water soluble at the prepared concentrations. Therefore, the solutions were not optically scattering. The absorption coefficients of all the prepared solutions were obtained with Equation (4), and the optical absorbance spectra were measured with a UV-Vis-NIR spectrophotometer (Cary 6000i, Varian, Walnut Creek, CA, USA) in transmission mode. Figure 3 displays the absorption coefficients of the stock solutions.

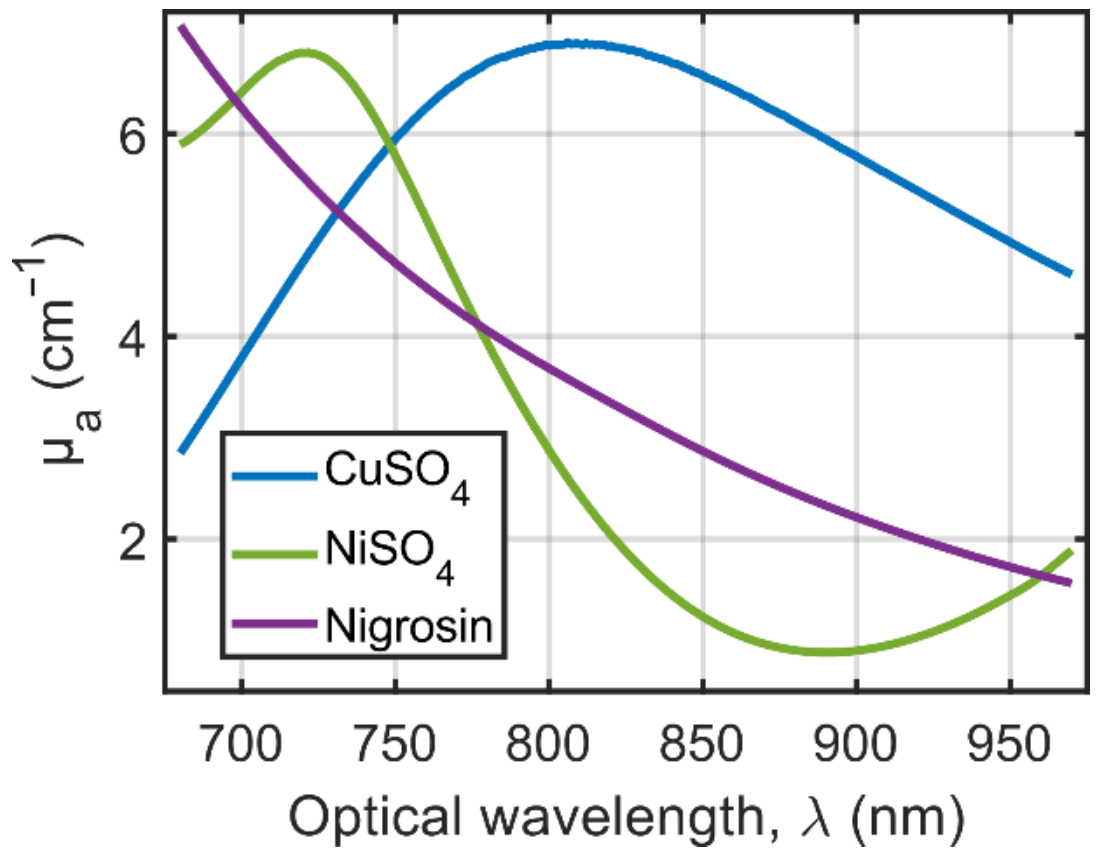

**Figure 3.** Optical absorption spectra of the stock solutions in the optical range 680–970 nm. The blank measurement was performed with ultrapure water.

### 2.3.2. Choice of the Calibration Solution

The calibration solution was chosen within the three non-fluorescent and non-scattering molecular solutions described in Section 2.3.1 so that $\delta_{tube}(\lambda)$ could be computed with Equation (6). Several key features were considered.

The first requirement for the calibration solution is the ability to determine its coefficient of absorption $\mu_a^{calibration}(\lambda)$ with SPP. All three solutions are non-scattering and hence meet this criterion. Second, the calibration solution should have a significant absorption coefficient over the entire investigated spectral range of 680–970 nm to ensure an accurate evaluation of $A^{PA}_{calibration}(\lambda)$. Figure 3 shows that none of the three compounds has a constant absorption coefficient over the considered range. However, $CuSO_4{\cdot}5H_2O$ can be considered to have the flattest spectrum because it has the lowest relative range of $\mu_a$ (ratio of the absolute difference between the maximum and minimum values to the mean value) (Table 2). Third, the calibration solution should be photostable and chemically stable so that the absorption spectrum does not change during the measurement process. Our investigations show that the three compounds meet those criteria. Moreover, inorganic compounds $CuSO_4{\cdot}5H_2O$ and $NiSO_4{\cdot}6H_2O$ were shown to have a long-term photostability (no photobleaching) even under exposure to high-power laser pulses [11]. For long-term use, the preparation of the calibration solution should be highly reproducible. The sulfate salts are available commercially with a high chemical purity, which ensures a good reproducibility. However, nigrosin is a mixture of organic dyes, so the variability from one batch to another should be considered. The last requirement is the determination of the PA generation efficiency (PGE) of the calibration solution $\eta_{calibration}$. The photothermal conversion efficiency $E_{pt}$ of the three stock solutions is assumed to be equal to 1, but the Grüneisen coefficient relative to water $\Gamma/\Gamma_{water}$ should also be known (Equation (2)). The molar absorptivity of nigrosin allows it to have a dye concentration around 1 mM for the stock solution. Therefore, the contribution of the dye to the Grüneisen coefficient can be neglected to assume that $\eta_{nigrosin}$ = 1. For $CuSO_4{\cdot}5H_2O$ and $NiSO_4{\cdot}6H_2O$, Fonseca et al. [11] reported that $\Gamma/\Gamma_{water}$ is wavelength-independent in the range 740 to 1100 nm. However, the molar concentration for the stock solutions was several orders of magnitude larger than for nigrosin (Table 2) and, at these concentrations of sulfate salts, the Grüneisen coefficient significantly differs from $\Gamma_{water}$ [11]. Fonseca et al. [11] proposed an empirical formula to determine $\Gamma/\Gamma_{water}$ of the sulfate solutions. However, we were not able to confirm this empirical formula.

With all these criteria and because cupric sulfate pentahydrate solutions have already been used as a model medium in PAI [11,22], our choice was: (1) to use the stock solution of $CuSO_4{\cdot}5H_2O$ as the calibration solution and (2) to determine $\eta_{calibration}$ using measurements performed with solutions of nigrosin.

#### 2.3.3. Commonly Used PA Contrast Agents

The PA spectrometer was tested on two different PA contrast agents based on nanoparticles and dyes, respectively. First, the nanoparticle solution was a commercial dispersion of citrate-capped gold nanorods (GNR) in water (10 ± 2 nm diameter, 42 ± 8 nm length, concentration 35 µg/mol, Sigma-Aldrich) with a nominal maximum extinction at 808 nm. The attenuation coefficient $\mu^{SPP}(\lambda)$ was evaluated with SPP (V650, Jasco, Japan) in the wavelength range 680–900 nm. SPP baseline correction and PAI blank datasets were performed with ultrapure water. Due to scattering, $\mu^{SPP}$ is expected to differ from $\mu_a$.

For the dye agent, solutions of indocyanine green (ICG, pharmaceutical primary standard, Sigma-Aldrich) at different concentrations were prepared. First, a stock solution was prepared by dissolving 7.4 mg of powder in 1.5 mL of dimethyl sulfoxide (DMSO) (concentration of 6.4 mM). This stock solution was diluted to obtain 5 concentrations of ICG: 5.5, 7, 9, 12, and 15 µM, each in 25.0 mL of solvent. The final solvent composition was 98.9% Dulbecco's Phosphate Buffered Saline (concentrated ×1 DPBS, Gibco), 1% DMSO, and 0.1% Tween 20 (Sigma-Aldrich). Tween 20 is a non-ionic surfactant that forms micelles and stabilizes the dye [23]. Additionally, a solution at 7 µM of ICG was prepared in a solvent without Tween 20. Since ICG is known to be unstable in aqueous solutions and photosensitive, the solutions were stored in amber glass vials, and PA spectra acquired within 1 h after their preparation. In parallel to the PAI measurements, the absorption coefficients of the solutions $\mu_a(\lambda)$ were measured by SPP (V650) in the wavelength range 680–900 nm. The scattering of the solution was negligible in the spectral range of interest. SPP baseline correction and PAI blank datasets were obtained with the solvent.

## 3. Results

### *3.1. Robustness of the Measurements with the Calibration Solution*

Robust measurements with the calibration solution are crucial for the reliability of the calibration process. Therefore, the calibration solution itself was used to evaluate the measurement variability.

#### 3.1.1. Measurement Repeatability

The measurement repeatability, under different injection conditions, was evaluated simultaneously in a series of 10 acquisitions. In tube 1 (Figure 1c), the calibration solution was injected before the first acquisition and left untouched for the whole series to assess the intrinsic measurement fluctuations of the system. For tube 2, 50 µL of the calibration solution was injected before each acquisition (without flushing with water and air) to measure the variations due to the injection process. Tube 3 was flushed with air only, and the calibration solution was injected between each acquisition. Flushing with air prevented the mixing of samples corresponding to two successive acquisitions. Finally, tube 4 was cleaned with water and air, and the calibration solution was injected between successive acquisitions.

For each tube, $\tilde{A}^{PA}_{calibration}(\lambda)$ was computed over the 10 acquisitions (Figure 4a). The values depend on the tube, and the variations can be attributed to the spatial heterogeneities of the light distribution and of the ultrasound detection. The wavelength dependency is caused by the spectrum of the calibration solution, the spectral sensitivity of the power meter, and the wavelength-dependent laser power at the fiber input and after propagation in the optical fibers and in water. Then, $\Psi^{PA}{}_i(\lambda)$ was computed from Equation (7) for each acquisition *i*. The expected value for $\Psi^{PA}{}_i(\lambda)$ is one. The median relative error MRE($\Psi^{PA}(\lambda)$,1) was calculated for each tube and each wavelength $\lambda$ according to equation (11) and is displayed in Figure 4b. Additionally, each spectrum $\Psi^{PA}{}_i$ was fitted to a flat spectrum (of amplitude one for all $\lambda$), assuming a direct proportionality. The fitting factor $\gamma_i$ has an expected value of one. The MRE of $\gamma_i$ for each tube is shown in Figure 4c.

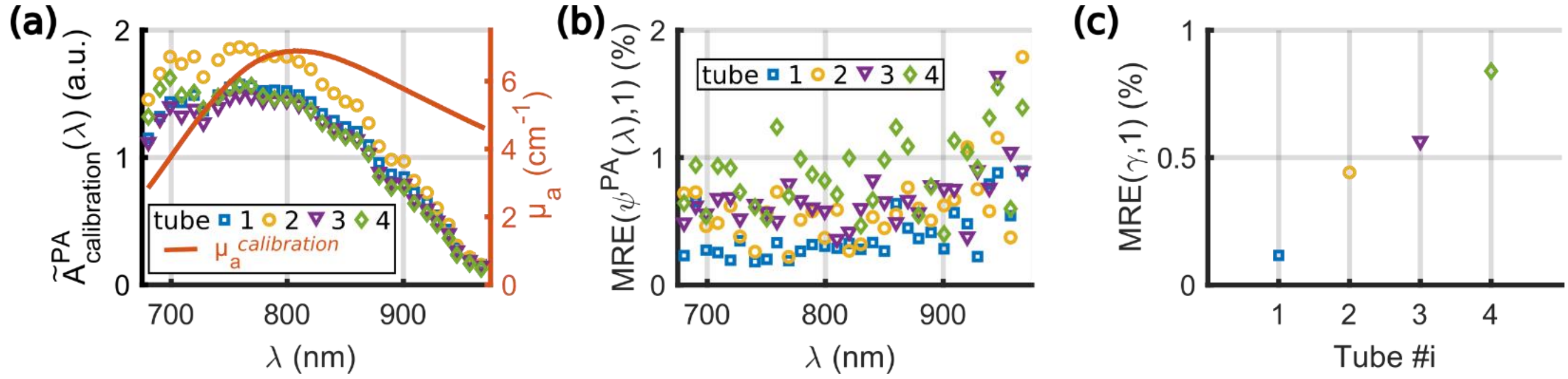

**Figure 4.** Repeatability of measurements with the calibration solution evaluated with 10 acquisitions. Different experimental conditions were applied for the 4 tubes. Tube 1: the solution was injected once. Tube 2: the solution was re-injected. Tube 3: the tube was flushed with air, and the solution was re-injected. Tube 4: the tube was flushed with air and water, and the solution was re-injected. (**a**) $\tilde{A}^{PA}_{calibration}(\lambda)$ used for the evaluation of $\Psi^{PA}$ for each tube and the absorption spectrum of the calibration solution. (**b**) Median relative error (MRE) of $\Psi^{PA}$ vs. the optical wavelength $\lambda$ for each tube. (**c**) MRE for the fitting factor of $\Psi^{PA}$ by a flat spectrum of amplitude unity.

The MRE for $\Psi^{PA_i}(\lambda)$ is below 2% for all the wavelengths and the tubes, while they are below 1% for the fitting factor $\gamma$. These low percentages demonstrate an excellent repeatability of the measurement. The main source of fluctuations is the injection of the solution. Flushing with air resulted in similar fluctuations as re-injecting without flushing. However, the injection with cleaning (tube 4) had the strongest variation. This variation could be attributed to droplets of the solvents, which may stay in the tube (or the needle) and could result in dilution of the injected solution. The global fluctuations in amplitude of the spectrum (Figure 4c) were found to be lower than for individual wavelengths (Figure 4b), which suggests additional sources of fluctuations at each wavelength. For all tubes, the MRE is stable over the wavelength range 680–920 nm and increases in the range 920–970 nm. This increase could be attributed to the lower laser fluence at the tube location above 920 nm. Indeed, the absorption of the laser radiation by water between the fiber output and the sample is stronger above 920 nm [24]. Consequently, $\tilde{A}^{PA}_{calibration}(\lambda)$ is smaller (Figure 4a), resulting in an amplification of the errors in the estimation of $\Psi^{PA_i}(\lambda)$ for $\lambda > 920$ nm.

#### 3.1.2. Influence of the Number of Acquisitions for the Evaluation of $\tilde{A}^{PA}_{calibration}(\lambda)$

The evaluation of $\tilde{A}^{PA}_{calibration}(\lambda)$ is crucial for the calibration of the system and needs to be performed for each series of measurements, since no precise alignment is executed. Because of the measurement variability, the number of acquisitions with the calibration solution required to have a robust estimate of $\tilde{A}^{PA}_{calibration}(\lambda)$ needs to be evaluated.

The following experiment was carried out. Seven pairs of acquisitions with the calibration solution injected in the four tubes were performed (total of 14 acquisitions). Between two successive pairs, at least 10 acquisitions with other solutions were performed. Each pair was preceded with a blank acquisition of water, and the four tubes were flushed with air only between the two injections of the calibration solution.

First, $\tilde{A}^{PA}_{calibration}(\lambda)$ was computed per tube by taking the 14 acquisitions. MRE($\Psi^{PA}(\lambda)$,1) is displayed in Figure 5a and shows values below 2%, similarly to Figure 4b. MRE($\gamma$,1) was equal to 1.2% when $\Psi^{PA}$ was fitted with a flat spectrum of amplitude unity. It should be notated that the 56 evaluations (14 acquisitions in four tubes) of $\Psi^{PA}(\lambda)$ and $\gamma$ were considered, adding the inter-tube variability compared to in Section 3.1.1.

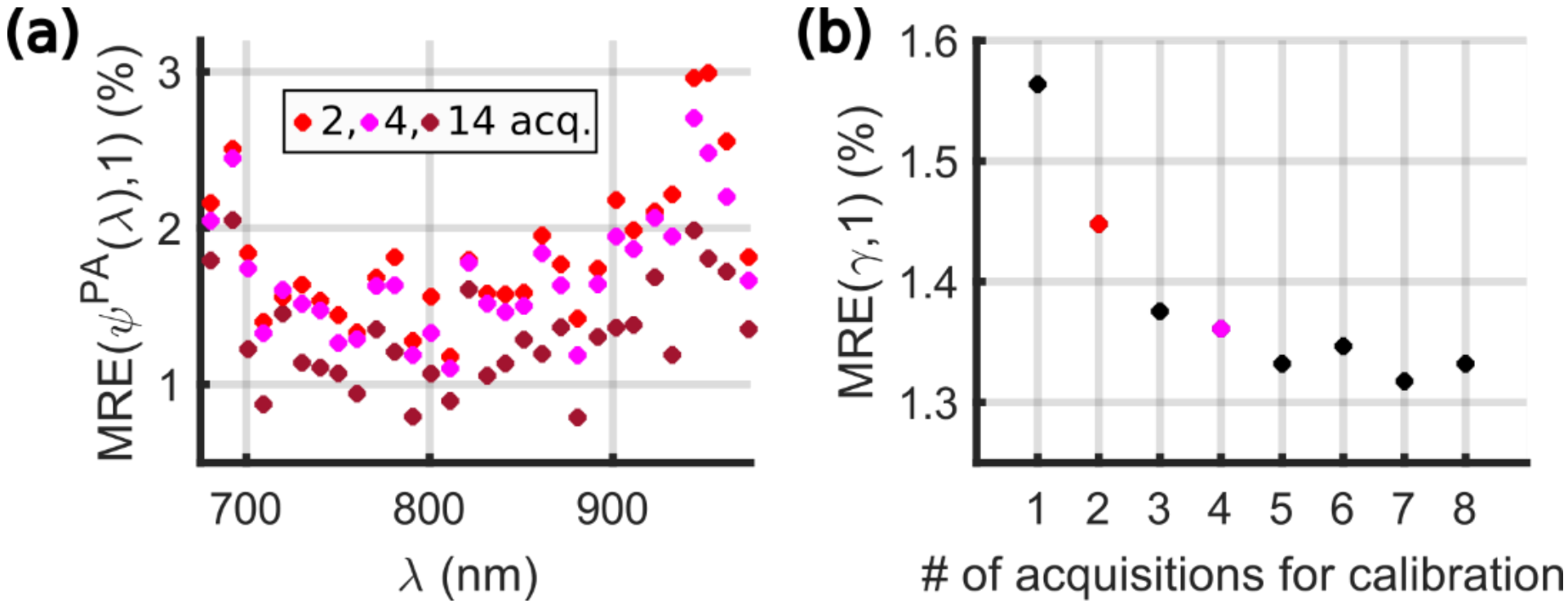

**Figure 5.** Evaluation of the influence of the number of acquisitions taken to compute the median of $A^{PA}_{calibration}(\lambda)$. (**a**) Median relative error (MRE) of $\Psi^{PA}$ vs. the optical wavelength $\lambda$ for combinations of 2, 4, and 14 acquisitions out of 14 to evaluate the median of $A^{PA}_{calibration}(\lambda)$ per tube. (**b**) MRE for the fitting factor for combinations from 1 to 8 acquisitions out of 14 to evaluate the median of $A^{PA}_{calibration}(\lambda)$ per tube.

Second, $\tilde{A}^{PA}_{calibration}(\lambda)$ was computed per tube by taking $m = 1$ to $m = 8$ acquisitions. Then, $\Psi^{PA}(\lambda)$ and $\gamma$ were computed for the $(14 - m)$ other acquisitions. All the combinations of 14 acquisitions taken $m$ at a time were used for the computation of the MRE, which corresponds to 4368 spectra for $m = 2$. Figure 5b shows the global decrease of the error with increasing $m$. The statistical evaluations with $m = 2$ and $m = 4$ are highlighted in Figure 5. The error is statically below 3% for the entire spectral range.

#### 3.1.3. Measurement Protocol for a Series of Samples

In practice, we chose to start and to end each series of measurements with two consecutive acquisitions of the calibration solution to have at least four acquisitions (two before and two after) per tube to compute $\tilde{A}^{PA}_{calibration}(\lambda)$. For the samples, the spectra were computed using two consecutive acquisitions. Between the two acquisitions of the same sample, the tubes were flushed with air before being injected again with 15 µL of the sample. Between two different samples, the tubes were flushed with air and the solvent to clean them and again with air to avoid dilution of the next sample. A blank dataset was acquired between two different samples when the tubes were filled with the solvent.

### *3.2. Characterization of the PA Spectrometer with Non-Fluoresent and Non-Scattering Molecular Solutions*

#### 3.2.1. Linearity of the PA Spectrometer

The linearity of the spectrometer was tested on the nigrosin solutions, as the PGE of nigrosin $\eta_{nigrosin}$ is expected to be independent of the concentration and the wavelength. Experiments were performed at 20.5 °C ± 0.2 °C (median ± MAD). $\tilde{A}^{PA}_{calibration}(\lambda)$ was computed from six acquisitions per tube. $\xi^{PA}$ was computed for the five percent solutions of nigrosin and for four acquisitions in four tubes for each solution. Figure 6a shows the proportionality of $\xi^{PA}$ to the absorption coefficient for three wavelengths and the five solutions. The proportionality factor should be equal to $1/\eta_{calibration}$ and was found to be equal to 0.79, which indicates that the PGE of the calibration solution $\eta_{calibration}$ is, as expected, greater than one. Figure 6b presents the measured spectra for the 40% and the 60% solutions. Only two solutions were presented here for the sake of legibility. The spectra $\xi^{PA}(\lambda)$ were fit with a corresponding absorption spectra $\mu_a(\lambda)$ assuming a direct proportionality. The error between the fitted spectrum and $\xi^{PA}(\lambda)$ was found below 5% for $\lambda < 930$ nm (Figure 6c). These results demonstrate that the spectral shape of the nigrosin solution could be retrieved by computing $\xi^{PA}(\lambda)$ using the calibration solution of $CuSO_4{\cdot}5H_2O$. The proportionally factor was assessed on the five solutions to be equal to 0.796 ± 0.015 (median ± MAD), which corresponds to $\eta_{calibration} = 1.256 \pm 0.023$.

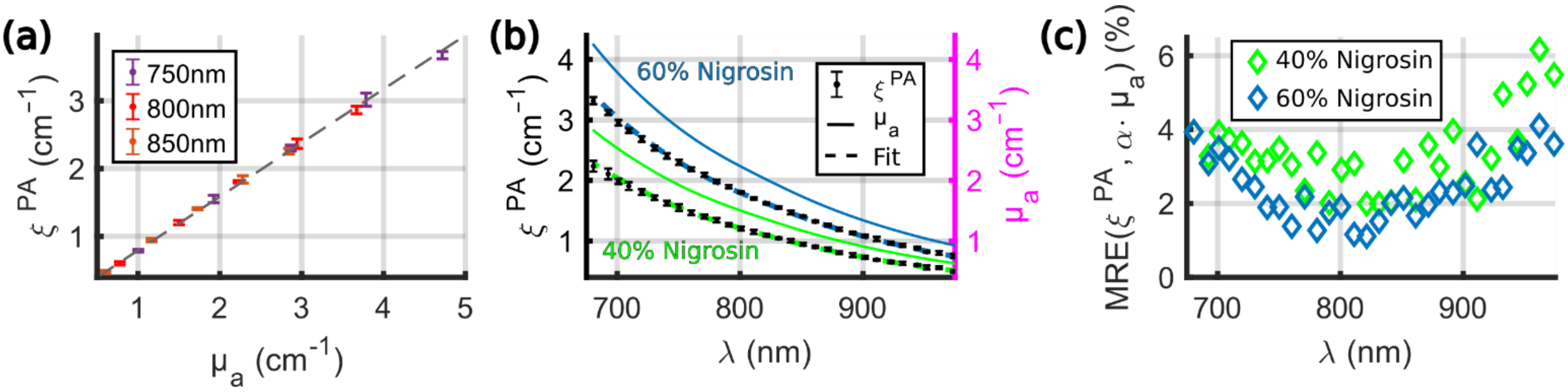

**Figure 6.** Linearity of the PA spectrometer with respect to the absorption coefficient $\mu_a$ and at different wavelengths. $\xi^{PA}$ was computed for the 5 percent solutions of nigrosin and for 4 acquisitions in 4 tubes for each solution. (**a**) $\xi^{PA}$ values (median ± MAD) as a function of $\mu_a$ for 3 different wavelengths and the 5 solutions. The dashed line is a fit by a homogeneous linear function of slope 0.79. (**b**) The left axis corresponds to $\xi^{PA}$ values (median ± MAD) as a function of the optical wavelength for two solutions: 40% and 60% of the stock solution. The absorption spectra $\mu_a$ are displayed (right axis, solid line), as well as the fits. The proportionality factors were $\alpha_{40\%} = 0.81$ and $\alpha_{60\%} = 0.80$, respectively. (**c**) MRE of $\xi^{PA}$ with $\alpha{\cdot}\mu_a$ as a reference for the 40% and 60% percent solutions.

#### 3.2.2. Evolution of $\eta_{sample}/\eta_{calibration}$ with the Concentration and the Temperature for the Different Solutions

For the percent solutions of $CuSO_4$ and $NiSO_4$, and for the solution mix-$SO_4$, the PGE was shown to be wavelength-independent in the range 740 nm to 980 nm [11]. For the percent solutions of nigrosin, the PGE is expected to be wavelength-independent for the range 680–980 nm. Hence, the spectra $\xi^{PA}(\lambda)$ are expected to be directly proportional to the absorption spectra $\mu_a(\lambda)$, in the range where the PGE is constant.

$$\xi^{PA}(\lambda) = \frac{\eta_{sample}}{\eta_{calibration}} \cdot \mu_a^{sample}(\lambda) = \alpha_{sample} \cdot \mu_a^{sample}(\lambda) \tag{12}$$

According to Fonseca et al. [11], for the solutions of $CuSO_4$ and $NiSO_4$, the coefficient $\alpha_{sample}$ depends on the concentration with the empirical law:

$$\alpha_{sample} = \alpha_0 \cdot (1 + \beta_{CuSO4} \cdot c_{CuSO4} + \beta_{NiSO4} \cdot c_{NiSO4}) \tag{13}$$

where $c_{CuSO4}$ and $c_{NiSO4}$ are the molar concentrations of $CuSO_4{\cdot}5H_2O$ and $NiSO_4{\cdot}6H_2O$, respectively, and theoretically, $\alpha_0 = 1/\eta_{calibration}$. The coefficients $\beta_{CuSO4}{}^{F} = 0.708\ M^{-1}$ (at 23.0 °C) and $\beta_{NiSO4}{}^{F} = 0.325\ M^{-1}$ (at 22.6 °C) were determined experimentally for wavelengths between 1400 and 1500 nm where the absorption was dominated by water and not by the solutes.

The PA spectra $\xi^{PA}(\lambda)$ of the solutions of $CuSO_4$, $NiSO_4$, and nigrosin, as well as the solution mix-$SO_4$, were measured at two different temperatures of the water bath: 20.3 °C ± 0.2 °C and 25.0 °C ± 0.2 °C. Samples with similar absorption were grouped together, and the groups were measured by ascending order of absorption. Each of the 16 samples was injected and measured twice in four tubes. The calibration solution was measured twice before and after each group, resulting in 12 measurements per tube to compute $\tilde{A}^{PA}_{calibration}(\lambda)$. The proportionality factors $\alpha_{sample}$ were evaluated with a curve-fitting algorithm from $\tilde{\xi}^{PA}(\lambda)$. Figure 7a–e presents the spectra $\xi^{PA}(\lambda)$ for the different groups of solutions, together with the fits and the measured spectra $\mu_a(\lambda)$. Figure 7f presents $\alpha_{sample}$ as a function of the dilution percentage of the stock solution. The solution mix-$SO_4$ was placed in Figure 7c (intitled 60%) due to the absorption values, but it was placed at 30% in Figure 7f because of the percentage of the stock solution of $NiSO_4$ used.

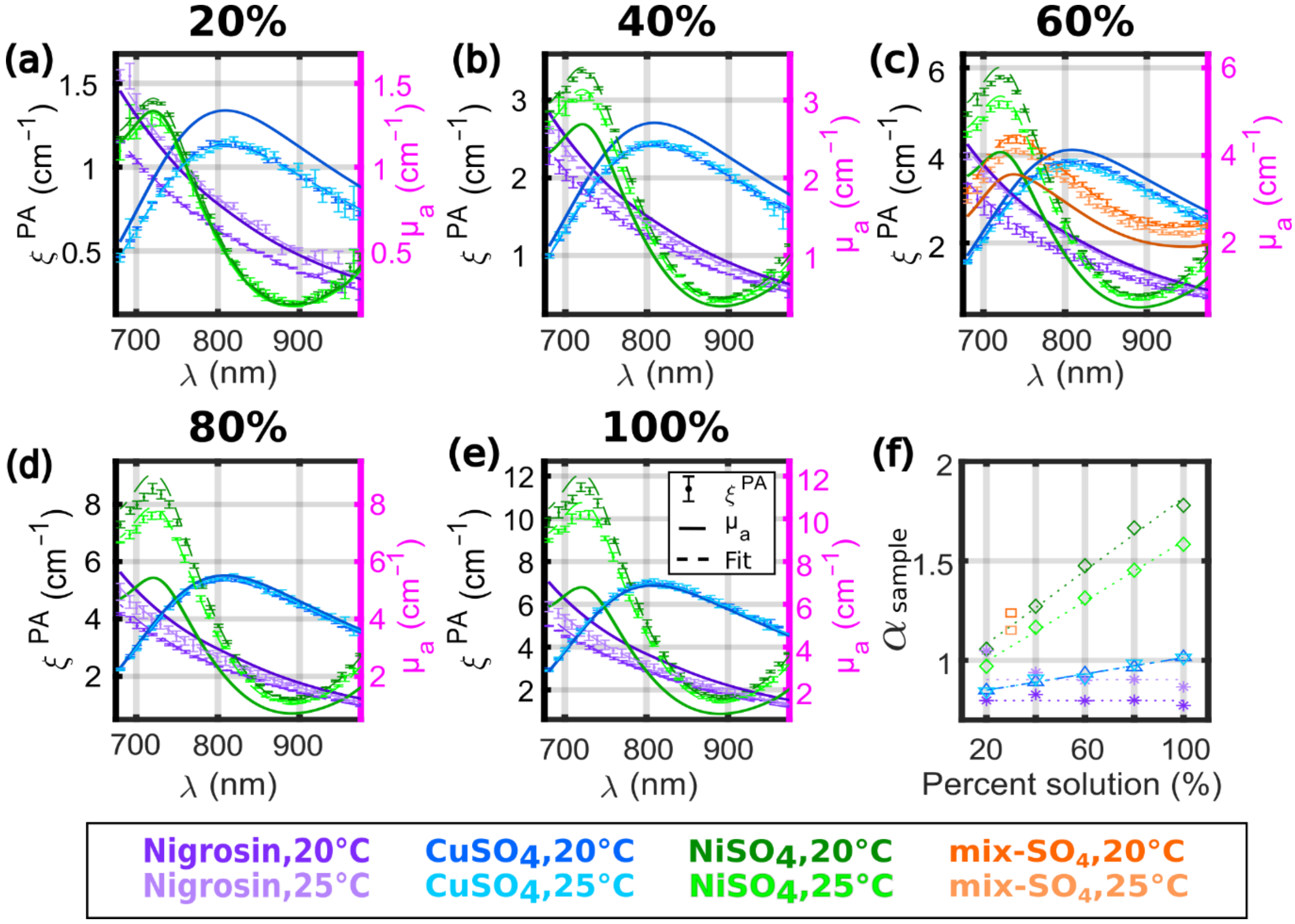

**Figure 7.** Influence of the concentration and the temperature on the spectra $\xi^{PA}(\lambda)$ for the nigrosin and the sulfate solutions. (**a**–**e**) Spectra for the different percent solutions and two different temperatures (20 °C and 25 °C). The absorption coefficient $\mu_a(\lambda)$ is displayed with a solid line. The spectra $\xi^{PA}(\lambda)$ were measured twice in four tubes (total of 8 evaluations) for each temperature, and the median ± MAD values were displayed with dot markers and error bars. Two different tones of the same color were used for the different temperatures, and the lighter tone was used for 25 °C. The fits of $\mu_a(\lambda)$ to $\tilde{\xi}^{PA}(\lambda)$ (with an assumption of direct proportionality) are shown with a dashed line of the same color as for $\xi^{PA}(\lambda)$. The percent solutions are separated in 5 graphs for the sake of legibility: (**a**) 20%, (**b**) 40%, (**c**) 60%, (**d**) 80%, and (**e**) 100%. The mix-$SO_4$ is displayed in (**c**). (**f**) The proportionality factor between $\tilde{\xi}^{PA}(\lambda)$ and $\mu_a(\lambda)$ as a function of the percent of the stock solution. The mix-$SO_4$ is displayed at 30% because it is comprised of 30% of the stock solution of $NiSO_4$. The dotted lines are the linear regressions. The same colors as for $\xi^{PA}(\lambda)$ were used.

For the solutions of nigrosin, the different spectra $\xi^{PA}(\lambda)$ were fitted by the corresponding spectra $\mu_a(\lambda)$ over the entire spectral range for all the percent solutions. However, the values of $\alpha_{sample}$ were larger at 25 °C than at 20 °C for all the samples. The values of $\alpha_{sample}$ could be considered as independent of the concentration, and the constant value was evaluated by the median (Table 3). The corresponding PGEs of the calibration solution were $\eta_{calibration}$(20 °C) = 1.253 ± 0.003 (median ± MAD) and $\eta_{calibration}$(25 °C) = 1.107 ± 0.062. The value at 20 °C is consistent with the value found with the experiment described in Section 3.2.1., while the two series of measurements were performed at different dates. With Equation (13) and the value $\beta_{CuSO4}^F$, the computed PGE at 23 °C will be $\eta^F_{calibration}$(23 °C) = 1.177. Therefore, the measured and computed evaluations of $\eta_{calibration}$ consistently decrease with increasing temperature. This result indicates that the temperature of the water bath is an important parameter to obtain an accurate quantification with our calibration method. The median value of $\eta_{calibration}$ determined with the nigrosin solutions was used for the computation of $\theta^{PA}(\lambda)$ with Equation (8).

For the solutions of $CuSO_4$, we can notice that the spectra $\xi^{PA}(\lambda)$ at 20 °C and at 25 °C are superimposed and are fitted by the corresponding spectra $\mu_a(\lambda)$ over the entire spectral range. As expected, $\alpha_{CuSO4,100\%}$ is close to one and increases linearly with the increasing concentration of $CuSO_4 \cdot 5H_2O$. $\beta_{CuSO4}$ was determined by linear regression to be of the order of 1.0 (Table 3), which is 41% higher than the value of $\beta_{CuSO4}^F$ determined by Fonseca et al. For $NiSO_4$, $\xi^{PA}(\lambda)$ has higher values at 20 °C than at 25 °C. For the stock solution (100%), the coefficient $\alpha_{sample}$ was found to be larger than 1.5, and the values of $\xi^{PA}(\lambda)$ were lower than $\alpha_{sample} \cdot \mu_a(\lambda)$ for $\lambda < 740$ nm. For the 20% and 40% solutions, the fit matches over the whole spectral range (even for $\lambda < 740$ nm). By fitting for $\lambda > 740$ nm, the coefficient $\alpha_{sample}$ was found to vary linearly with the concentration, but the coefficient $\beta_{NiSO4}$ was found to be equal to 0.74 $M^{-1}$ at 20 °C and 0.65 $M^{-1}$ at 25 °C. These values of $\beta_{NiSO4}$ are at least two times larger than the values determined by Fonseca et al. The values of $\alpha_0$ (Table 3) also differ from the empirical model proposed by Fonseca et al. Indeed, they were determined to be in the same order of magnitude as for the nigrosin, but they did not match the values or the variation with the temperature found for the nigrosin. $\alpha_0$ was found to be constant for $CuSO_4$ and decreasing for $NiSO_4$.

For the mix-$SO_4$, we found $\alpha_{mix\text{-}SO4}$(20 °C) = 1.24 and $\alpha_{mix\text{-}SO4}$(25 °C) = 1.15, while the value from Fonseca et al. would be $\alpha^F_{mix\text{-}SO4}$(23 °C) = 1.01. While each of the stock solutions contributes to 30% of the mix-$SO_4$, the molar concentration of $NiSO_4$ in the solution is about five times larger than $CuSO_4$. Thereby, the contribution of $NiSO_4$ dominates in $\alpha_{mix\text{-}SO4}$, which explains the values above one and the decreasing value with the increasing temperature. However, because of the various values of $\alpha_0$ for $CuSO_4$ and $NiSO_4$, we were not able to verify an empirical law similar to Equation (13). The discrepancy between our results and the results of Fonseca et al. are further discussed in Section 4.

**Table 3.** Coefficients of Equation (13) obtained by linear regression of the curves $\alpha_{sample}$ as a function of the molar concentration of each compound. The coefficient of determination ($R^2$) was computed.

| Sample | Temperature (°C) | $\alpha_0$ | $\beta$ ($M^{-1}$) | $R^2$ |
|---|---|---|---|---|
| CuSO4 | 20.3 | 0.81 | 1.00 | 0.99 |
| | 25.0 | 0.80 | 1.01 | 0.95 |
| NiSO4 | 20.3 | 0.90 | 0.74 | 0.99 |
| | 25.0 | 0.84 | 0.65 | 0.99 |
| Nigrosin * | 20.3 | 0.80 | - | - |
| | 25.0 | 0.90 | - | - |

* The coefficient $\alpha_0$ for the nigrosin is the median of the $\alpha_{sample}$ for the percent solutions.

### 3.3. Characterization of Conventional Contrast Agents

#### 3.3.1. Gold Nanorods

The solution of gold nanorods (GNR) was measured in a series of acquisitions together with the 20% solutions of nigrosin and $CuSO_4$. The experiments were performed at 20.7 ± 0.2 °C, and the $\eta_{calibration}$ determined in Section 3.2.1 was used to compute the PA coefficient $\theta^{PA}(\lambda)$. The laser fluence was lowered at 2 mJ.cm$^{-2}$ at 730 nm with polarizing optics placed before the input of the fiber bundle. The lower fluence aimed at preserving the photostability of the gold nanorods. Figure 8a displays the spectra $\theta^{PA}(\lambda)$ for the solutions and the attenuation coefficient $\mu^{SPP}(\lambda)$ measured by SPP, and Figure 8b presents the ratio between the two quantities in the range 680–900 nm. For the non-scattering solutions, (nigrosin and $CuSO_4$), $\mu^{SPP}(\lambda)$ is equal to the absorption coefficient $\mu_a(\lambda)$, and the ratio $\tilde{\theta}^{PA}/\mu_a$ was found to be constant, as expected from Section 3.2.2. The median of the ratio was found to be equal to 0.99 for the nigrosin, which validates the capabilities of the system to operate at different laser fluences. For $CuSO_4$, the PGE was estimated to 1.08, as in Section 3.2.2.

For the GNR, we first verified that the spectra of the sample were stable during the measurement by computing the spectra $A^{PA}$ without averaging the ultrasound data over the 15 laser sweeps. Then, $\theta^{PA}(\lambda)$ was computed with the averaged US data. $\theta^{PA}(\lambda)$ and $\mu^{SPP}(\lambda)$ had a similar spectral shape: a broad peak centered around 800 nm and a full width at half the maximum of ~130 nm. However, the spectral shapes did not match with a direct proportionality factor (proportionality factor for the best fit estimated to 0.74). The ratio $\tilde{\theta}^{PA}/\mu^{SPP}$ was found to be approximately equal to one from 680 nm to 710 nm and then decreasing until 800 nm and quasi constant from 800 nm to 900 nm. This ratio is not the PGE because the attenuation coefficient of GNR comprises both the absorption and the scattering coefficients. The lower value of $\theta^{PA}$ compared to $\mu^{SPP}$ could be explained by the combination of two phenomena. First, the scattering for the solution of GNR cannot be neglected, which would result in an overestimation of $\mu^{SPP}$ as compared to $\mu_a$. For the geometrical characteristics of the GNR used here, the shapes of the absorption and scattering spectra are expected to match (but could be slightly shifted), and the ratio between the scattering and the absorption coefficients could be on the order of 10% [25]. Another phenomenon must be accounted for to describe the discrepancy: the GNR is the photothermal converter (heat source) while water is the PA signal-generating medium. Interfacial thermal resistance to the heat transfer at the gold–water interface could lower the effective photothermal conversion efficiency [26,27] and then reduce the value of $\theta^{PA}$. However, the influence of the interfacial thermal resistance on the PA signal is not well understood even at a single wavelength (peak absorption) [28], and by extension its spectral dependency is not yet established.

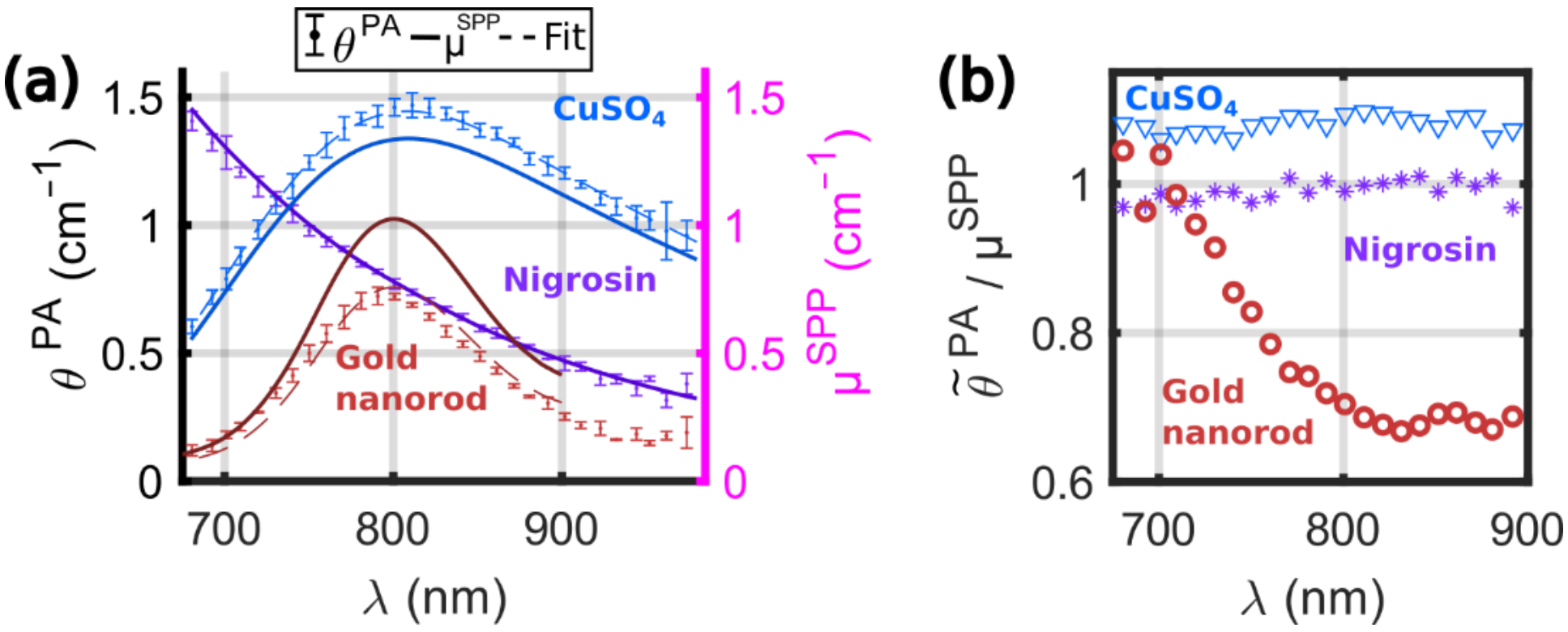

**Figure 8.** Experimental spectrum from a solution of gold nanorods (GNR). (**a**) PA coefficient $\theta^{PA}(\lambda)$ of the GNR solution (red curves), the 20% solutions of CuSO4 (blue curves), and the 20% solutions of nigrosin (purple curves) were measured once in four tubes. The median ± MAD values of $\theta^{PA}(\lambda)$ were displayed with dot markers and error bars. The attenuation coefficient $\mu^{SPP}(\lambda)$ is displayed with a solid line. For the GNR, the attenuation coefficient was only available in the range 680 nm–900 nm. The fits of $\mu^{SPP}$ to $\tilde{\theta}^{PA}$ are shown with a dashed line of the same color as for $\theta^{PA}(\lambda)$. (**b**) Ratios $\tilde{\theta}^{PA}/\mu^{SPP}$ for the different solutions as a function of the optical wavelength. The same colors as for $\theta^{PA}(\lambda)$ were used.

### 3.3.2. Indocyanine Green

Figure 9a displays images at $\lambda_2$ = 790 nm of a tube filled with (from left to right): the calibration solution, ICG in an aqueous solution without Tween, and ICG stabilized in micelles using 0.1% of Tween. For the calibration solution and the ICG with Tween, the images of the tube are similar: one spot with the same center and the same width. Therefore, we can assume that the calibration and the sample measurements match. However, for the ICG in an aqueous solution without Tween, the image of the tube has two spots along the depth dimension. The position of the dip between the two spots was found to correspond to the maximum of the single spot for the other solutions. Because of its affinity with hydrophobic surfaces, we can assume that ICG did not stay in the aqueous solution and stuck to the wall of the PTFE tube when Tween was not added. The two spots would then correspond to the interferences of the signals generated by the absorbing walls in the limited-view detection geometry. The calibration cannot be used for quantitative measurements when two spots appear because the amplitude $A^{PA}$ does not correspond to the same experimental conditions. These results indicate that surfactants or other compounds are needed to keep hydrophobic absorbers in solution for quantitative accuracy of the measurements. All the other measurements with ICG were performed in a solution comprised of 0.1% of Tween 20 [29].

The solutions of ICG were measured at a temperature of 22.0 °C ± 0.7 °C. $\eta_{calibration}$ was computed assuming a linear variation between 20 °C and 25 °C. A total of 12 evaluations were performed per solution. Figure 9b,c displays the spectra $\theta^{PA}(\lambda)$ and $\mu_a(\lambda)$, respectively, for five concentrations of ICG and in the range 680–900 nm. The absorption of ICG above 900 nm is negligible because of the shape of the spectra and was not displayed for better legibility. The scattering could be neglected, and $\mu_a(\lambda)$ was measured with SPP for the solutions of ICG. The PA and optical spectra have a peak around $\lambda_2$ = 790 nm and a shoulder around $\lambda_1$ = 742 nm. The PGE $\eta_{ICG}$ was assessed with Equation (3) and depends on both the wavelength and the concentration of ICG. Here, the solvent is composed of DPBS with 1% DMSO; therefore, the Grüneisen coefficient of the solvent is expected to be slightly larger than $\Gamma_{water}$. However, given its low concentration, ICG is not expected to influence the Grüneisen coefficient of the solution contrary to the sulfate salts. Yet, the optical properties of ICG can vary with the concentration [12]. At $\lambda_1$ and $\lambda_2$, it can be seen that $\eta_{ICG}$ increases with the concentration. The slope is larger at $\lambda_2$, but the $\eta_{ICG}$ values are larger at $\lambda_1$ than at $\lambda_2$ (Figure 9d). In a similar solvent but without Tween, Fuenzalida Werner et al. [12] determined that $\eta_{ICG}(\lambda_1) \approx 0.80$ and $\eta_{ICG}(\lambda_2) \approx 0.62$ at concentrations below 10 µM. These results are consistent with our measurements at $c_{ICG}$ = 9 µM.

Figure 9f displays the spectra $\mu_a(\lambda)$ normalized by the ICG concentration. The amplitude of the shoulder at $\lambda_1$ was found to be linear with the concentration with a slope of one, while $\mu_a(\lambda_2)/c_{ICG}$ decreased with the increasing concentration because of the aggregation of ICG at higher concentrations [12]. Therefore, the ratio $\mu_a(\lambda_1)/\mu_a(\lambda_2)$ increases with the increasing $c_{ICG}$ (Figure 9g). Interestingly, the ratio $\theta^{PA}(\lambda_1)/\theta^{PA}(\lambda_2)$ was found to be constant, suggesting a stability of the heat transfer from the molecule to the solvent regardless of the concentration and, by extension, of the aggregation state. Figure 9e shows that the spectral shape of $\theta^{PA}(\lambda, c_{ICG})$ does not depend on the concentration of ICG. This stability of the spectral shape was not observed in water without Tween [12], probably because of a stronger aggregation of ICG molecules when they are not in micelles of Tween. The normalization factor used for $\theta^{PA}(\lambda, c_{ICG})$, however, is not directly $1/c_{ICG}$ as for $\mu_a(\lambda)$, but it is proportional to $1/(\eta_{ICG}(\lambda_1, c_{ICG}) \times c_{ICG})$. Therefore, the PA signal generation increases nonlinearly with $c_{ICG}$. The aggregation of ICG is expected to decrease the fluorescence efficiency, a competitive process to PA, and therefore benefits PA generation at higher concentrations, which is seen by the increase of $\eta_{ICG}$ (Figure 9d).

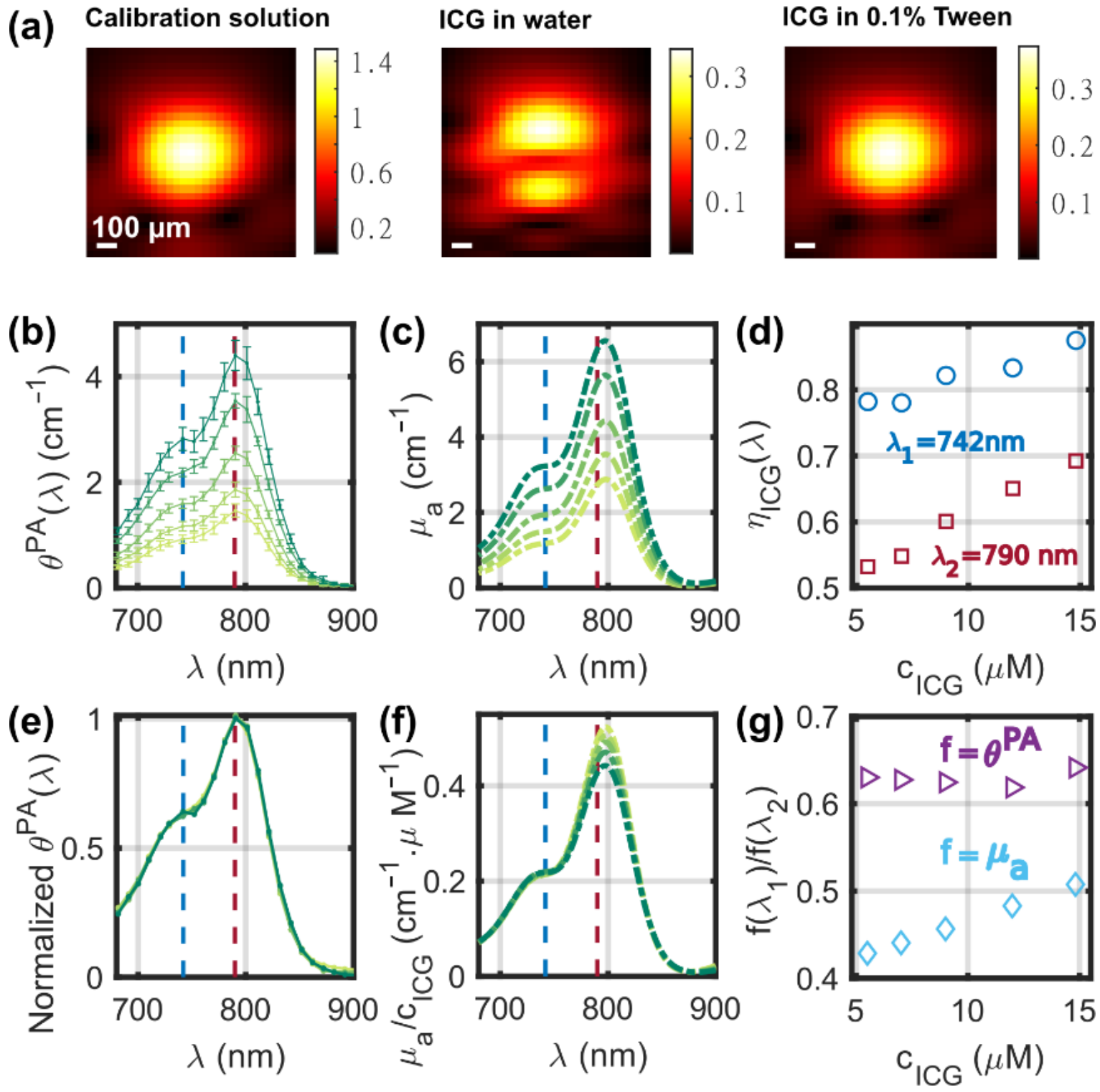

**Figure 9.** Experimental results for the solutions of indocyanine green (ICG). (**a**) PA images at 790 nm of a tube filled successively with: (left) the calibration solution, (middle) a solution of ICG in DPBS with 1% DMSO, and (right) a solution of ICG in DPBS with 1% DMSO and 0.1% Tween 20. The color scales are in arbitrary units. (**b**) PA coefficient $\theta^{PA}$ (median ± MAD) and (**c**) the corresponding $\mu_a$ spectra for the different concentrations of ICG. (**d**) Photoacoustic generation efficiency evaluated at two wavelengths, $\lambda_1$ = 742 nm and $\lambda_2$ = 790 nm, as a function of the concentration of ICG. (**e**) Median PA spectra of (**b**) normalized to their maximum value. (**f**) $\mu_a$ spectra of (**c**) normalized by concentration in ICG. For (**b–f**), the color of the curves changes from light green to dark green with increasing concentration. (**g**) Ratios $\theta^{PA}(\lambda_1)/\theta^{PA}(\lambda_2)$ and $\mu_a(\lambda_1)/\mu_a(\lambda_2)$ as a function of the concentration of ICG.

## 4. Discussion

We presented and validated a method to perform calibrated photoacoustic spectrometry in the wavelength range 680 to 970 nm with a commonly used PAI setup. The method requires successive injections in tubes that remain mechanically fixed and immersed in a water bath at a stable temperature. Three solutions need to be injected: water or a solvent as a background reference, a solution of $CuSO_4{\cdot}5H_2O$ at 0.25 M as a calibration solution, and the sample of interest. The simple calibration process provides PA spectra in spectroscopic units that can be related to the optical attenuation spectra (SPP), both in terms of shape and amplitude. The ratio between the PA spectrum $\theta^{PA}$ and the optical absorption spectrum $\mu_a$ yields the photoacoustic generation efficiency of the sample. Even if the absorption spectrum is not available for some samples, for instance, because of strong scattering, the PA spectra expressed in SI units can be compared to other samples.

The developed method requires the sample containers, the illumination, and the ultrasound detector to be stable during a series of measurement, but their respective position does not require a precise alignment. The method can even operate with a steady but nonuniform fluence distribution and ultrasound sensitivity. We also showed that the method can operate when the tubes are embedded in a tissue phantom (Appendix B). The developed calibration method can be adapted to various PAI systems, as long as they provide access to the PA-generated ultrasound signals or the beamformed images prior to envelope detection (to allow coherent subtraction of the background) and the pulse-energy fluctuations of the excitation light at each wavelength. Although radio-frequency data may not be available on all ultrasound machines [30], they are usually accessible on current conventional PAI systems. We applied here the method to a PAI system based on a clinical ultrasound array and carefully adapted the sample size to the frequency bandwidth of the detector. For other systems, the tube radius may be adapted for optimal sensitivity using Equation (A2) in Appendix A. To ensure a good transmission of ultrasound waves, the wall thickness should be as small as possible compared to the acoustic wavelength in the material at the center frequency of the detector. Furthermore, the material of the tube should have an acoustic impedance close to water ($Z^{water}$ = 1.49 Mray), as in the case of PTFE ($Z^{PTFE}$ = 2.97 Mray [31]). The wavelength at the center frequency is 0.3 mm in the material ($v_s^{PTFE}$ = 1390 m.s$^{-1}$). With a wall thickness of 0.1 mm, the transmission coefficient T in pressure through the three-layered system water–PTFE–water is [32]: T = 75% at 25 °C and at 5 MHz. For comparison, a glass tube with the same wall thickness would yield T = 10%. Additional properties of the tube should be a weak optical absorption and scattering, and the material of the tube should be chosen to be chemically inert to avoid interaction with the sample. Our PTFE tubes had all the required properties, and they are commercially available in various diameters and wall thicknesses. For independent measurements using multiple tubes, the necessary spacing between them in the tube phantom should be at least superior to the spatial resolution of the system, and a margin should be taken to avoid artifacts caused by side lobes. For PAI systems with an array focused in the elevational direction, the tube may be placed in the acoustic elevational focus for higher sensitivity.

Using tubes as sample containers has several advantages over the containers used in other PA spectrometers: Beard et al. [9] used a homemade cuvette, Fuenzalida Werner et al. [12] used a single channel microscopy chip, and Pelivanov et al. [13] enclosed their sample within a diaphragm between two quartz plates. For these containers, only one sample can be measured at a time, and changing to another sample may require tedious preparation procedures. For our system, several tubes could be positioned in the imaged region. This allowed parallel measurements of different samples or for the evaluation of the variability of the measurement for one type of sample. Because tubes have two opened ends, they could also easily be flushed, and the same tube could be used for a series of successive measurements. As the PTFE tubes are cost-effective, they were replaced as soon as they were polluted (persistent and additional absorption compared to the first blank dataset) or degraded. The small volume of the tubes (15 µL) is an asset to test samples at very early stages of the development of new contrast agents and for screening when only small quantities have been synthesized. However, the injection challenges compared to large cuvettes induce a variability that was reduced here by performing statistical evaluations over several injections. An automated sample injector, as is used in high-performance liquid chromatography (HPLC), might be considered in future development.

The sensitivity of the spectrophotometer increases with the laser fluence and the detector sensitivity. The detection limit was not determined here, but Figures 7a and 8a show that the 20% solution of nigrosin was measured with a high sensitivity even with a low absorption coefficient of $\mu_a$ = 0.3 cm$^{-1}$ at the two incident fluences 4 mJ.cm$^{-2}$ and 2 mJ.cm$^{-2}$ (evaluated at 730 nm). For comparison, other PA spectrometers reported a lower bound of calibrated detection corresponding to an absorption coefficient $\mu_a$ of 5 cm$^{-1}$ for the system of Beard et al. [10], 0.4 cm$^{-1}$ for the system of Fuenzalida Werner et al. [12], and 1 cm$^{-1}$ for the system of Pelivanov et al. [13].

For the upper bound of calibrated detection, the absorption coefficient of the stock solution was limited to $\mu_{a\ \text{stock}} \leq$ 7 cm$^{-1}$, which corresponds to an absorption length of $\ell_{\text{stock}} = 1/\mu_{a\ \text{stock}} \approx$ 1.4 mm. As the tube radius is 14 times smaller than $\ell_{\text{stock}}$, the tubes could be considered as optically thin. Thereby, the decay of the optical fluence inside the tube lumen caused by the sample absorption and its influence on the amplitude of the PA signal can be neglected. Indeed, while the absorption coefficient of the nigrosin solution varies significantly (relative range of 1.5, Table 2) over the covered spectral range, the PA spectra were found to have the same spectral shape as the absorption spectrum, without any reduction of the signal at larger absorption coefficients (Figure 6). Moreover, the proportionality factor between $\tilde{\xi}^{PA}(\lambda)$ and $\mu_a(\lambda)$ (Figure 7f) was found to be constant even when the percentage of the stock solution varied from 20% to 100%. This latter result also demonstrates that only one calibration solution can be used for an optical absorption coefficient below 7 $cm^{-1}$. Pelivanov et al. [13] reported the use of several calibration solutions of $CuSO_4{\cdot}5H_2O$ to cover the absorption range 1–26 $cm^{-1}$, but the concentration of the calibration solution was not given. Fuenzalida Werner et al. [12] reported measurements with a maximum absorption coefficient of 5 $cm^{-1}$, while other PA spectrometers reported a much larger upper bound of detectability: 250 $cm^{-1}$ for the system of Beard et al. [10] and 26 $cm^{-1}$ to up to 300 $cm^{-1}$ for the system of Pelivanov et al. [13]. As contrast agents for biomedical photoacoustic imaging can easily be diluted, the absorption coefficient of the tested solution can be lowered to fit below an upper limit of calibrated detection of our system.

Our calibration method relies on the injection in the sample container of an absorbing solution ($CuSO_4{\cdot}5H_2O$) whose photoacoustic properties are known. Compared to India ink, used elsewhere as a reference solution [12], $CuSO_4{\cdot}5H_2O$ is a molecular absorber of small molecular weight and not a particle-based absorber, so the heat transfer to the solvent is direct, resulting in a photothermal conversion efficiency of one. Moreover, the solution is homogeneous (no sedimentation), even for volumes as small as 15 μL, and stable. Unfortunately, its Grüneisen coefficient relative to water depends on both the concentration and the temperature of the solution, and its values were previously reported in only one study, to our knowledge [11]. At the concentration used for the calibration solution, our estimation of the photoacoustic efficiency matches with the value reported by Fonseca et al. [11]. Pelivanov et al. [13] also used $CuSO_4{\cdot}5H_2O$ as a calibration solution, but they assumed $\eta_{CuSO4}$ = 1 even at concentrations up to 1 M; therefore, their calibration did not account for the increase in the Grüneisen coefficient with the concentration. This discrepancy was not detected probably because of their use of different concentrations of $CuSO_4{\cdot}5H_2O$ as calibration solutions and a lack of validation with another reference solution such as nigrosin. Fuenzalida Werner et al. [12] separated the correction method used to obtain the shape of the PA spectrum from the calibration method to estimate the amplitude of the spectrum. They developed a complex correction method adapted to their measurement setup and validated its ability to retrieve the shape of the absorption spectrum in the range 400 nm to 900 nm with solutions of $NiCl_2$. However, amplitude calibration was performed in the visible range at 570 nm with Brilliant Black BN (BBN), a dye that is photo-stable, non-florescent, and for which the photoacoustic efficiency is expected to be equal to $\eta_{BBN}$ = 1. We identified nigrosin as a dye with similar properties in the NIR, and we used this highly absorbing dye to evaluate the Grüneisen coefficient of our calibration solution of $CuSO_4$. With the determined Grüneisen coefficient, the spectral shape and the amplitude of unknown samples were simultaneously obtained.

The Grüneisen coefficients relative to water of solutions of $CuSO_4{\cdot}5H_2O$ and $NiSO_4{\cdot}6H_2O$ were measured by Fonseca et al. [11] for five different concentrations from 0.125 M to 1 M for $CuSO_4$ and 0.275 M to 2.2 M for $NiSO_4$. According to their study, the Grüneisen coefficients of our calibration solution (0.25 M of $CuSO_4$) and of the 40% solution of $NiSO_4$ (0.55 M of $NiSO_4$) should have the same value, while we found a Grüneisen coefficient at least 16% larger for the 40% solution of $NiSO_4$. At a given temperature close to 23 °C, they determined a linear variation of the Grüneisen coefficient with the concentration in each solute (Equation (13)). We also found a linear variation for concentrations from 0.05 M to 0.25 M for $CuSO_4$ and from 0.26 M to 1.3 M for $NiSO_4$. However, the $\beta$ factors (Equation (13)) had a large discrepancy (Table 3). For $CuSO_4$, our evaluation of $\beta_{CuSO4}$ did not depend on the temperature and was 41% larger than the value reported by Fonseca et al. ($\beta^F_{CuSO4}$). For $NiSO_4$, $\beta_{NiSO4}$ was found to be larger at low temperature and at least 100% larger than $\beta^F_{NiSO4}$. Additionally, we were not able to confirm the empirical formula for the mix-$SO_4$. One possible explanation for the discrepancy of the $\beta$ values could be the difference in the investigated concentration range and the simplistic linear assumption over the whole concentration range investigated by Fonseca et al. However, we could not investigate the same concentration range due to the difference in calibrated detection limits of the PA spectrometers. Another major difference between the measurement methods is the spectral range in which the Grüneisen coefficient was evaluated. Fonseca et al. measured the Grüneisen coefficient based on the absorption of water in a spectral range (1400–1500 nm) where the absorption coefficient of $CuSO_4$ and $NiSO_4$ can be considered negligible compared to water, whereas we measured the Grüneisen coefficient in a wavelength range where the solute has a strong absorption. We can hypothesize that the Grüneisen coefficient differs when the absorption coefficient of the solution is dominated by the solute instead of the solvent. Using the same measurement system as Fonseca et al., Stahl [10] evaluated the Grüneisen coefficient of an aqueous solution of $CuCl_2$ ($c_{CuCl2}$ = 200 $g.L^{-1}$) as a function of the wavelength from 750 nm to 1500 nm. At 20 °C, it was observed that the Grüneisen coefficient was stable up to 1150 nm when the absorption was dominated by the solute, while for wavelengths greater than 1380 nm, where the absorption was dominated by water, the Grüneisen coefficient increased by 14%. A possible, but not confirmed, explanation could be a Grüneisen coefficient specific to the hydration shells around the metallic ions of $Cu^{2+}$ compared to the rest of the bulk solution. The bulk solution is mainly excited when the absorption is dominated by the solvent, as is the case for the experiments reported by Fonseca et al., while the hydration shells would have a stronger influence when the absorption is dominated by the solute. Unfortunately, no study was reported for different concentrations of $Cu^{2+}$ and for $Ni^{2+}$. These results show that the Grüneisen coefficients of the sulfate solutions of $CuSO_4$ and $NiSO_4$ depend on the concentration, the temperature, and the wavelength range of excitation. Further studies would be needed to use these chromophores at different concentrations in phantoms for quantitative multiwavelength photoacoustic imaging. Such studies are beyond the scope of this paper.

Spectral measurements of commonly used contrast agents in PAI (gold nanorods and ICG) showed results compatible with previously reported studies and demonstrated the ability of the spectrometer to characterize different kinds of agents: metallic nanostructures and dyes. In particular, the wavelength and concentration-dependent photoacoustic generation efficiencies of ICG were verified. Variations of $\eta_{ICG}$ with the concentration are linked to the changes in the photothermal efficiency, in particular due to the dye aggregation and reduced fluorescence. For scattering solutions of gold nanorods, as already shown by Pelivanov et al. [13], the PA spectrum enables removal of the influence of the light scattering to access to the absorption properties of the solution. Our PA spectral measurement can be used to quantitatively characterize the PA properties of plasmonic nanoparticles and is expected to capture the shape of the absorption spectrum. However, the photothermal conversion efficiency may be lower than one due to thermal resistance from the absorbers to the solvent [26]. Therefore, our calibrated measurement may not exactly match the amplitude of the optical absorption coefficient. Comparison with the attenuation coefficient measured with SPP in transmission mode could provide information about the strength of the scattering and its spectral dependency.

## 5. Conclusions

The design of photoacoustic contrast agents has been demonstrated to be challenging, in particular because the photoacoustic spectrum may differ from the optical attenuation spectrum due to scattering and other physical processes. Therefore, a calibrated measurement of the photoacoustic spectrum and the photoacoustic generation efficiency is highly desirable at all stages of the development of PA contrast agents. We demonstrated a novel method that can be adapted to most commonly used photoacoustic imaging systems to obtain calibrated photoacoustic measurements in the NIR range. Measurements were performed with small sample volumes of 15 µL, and the detection sensitivity is lower than 0.3 $cm^{-1}$. The system enables PA measurements at very early stages of the development of new contrast agents. This method can benefit the material science and biomedical communities and satisfy the growing need for the characterization of photoacoustic contrast agents.

**Author Contributions:** Conceptualization, J.G. and F.G.; methodology, J.G. and M.S.; software, T.L. and M.S.; validation, T.L., M.S. and J.G.; formal analysis, J.G. and T.L.; investigation, T.L., M.S., Y.A. and C.L.; resources, G.R., F.G. and J.G; data curation, T.L.; writing—original draft preparation, J.G.; writing—review and editing, T.L., M.S., C.L., F.G. and G.R.; supervision, J.G.; project administration, J.G.; funding acquisition, J.G., F.G. and G.R. All authors have read and agreed to the published version of the manuscript.

**Funding:** This work project has received financial support from the CNRS through the MITI interdisciplinary programs (Defi Imag'IN, 80 Prime), Gefluc Paris, Ile-de-France, Emergence Sorbonne Université 2019–2020, France Life Imaging (ANR-11-INBS-0006), and from the European Union's Horizon 2020 research and innovation program under grant agreements no. 801305 (NanoTBTech) and N°68335 (FOLSMART). M. Sarkar acknowledges support from the Paris Region (Ile-de-France) under the Blaise Pascal International Chairs of Excellence. Imaging was performed at the Life Imaging Facility of Paris Descartes University (Plateforme Imageries du Vivant—PIV).

**Acknowledgments:** The authors thank Lise Abiven at the Laboratoire de la Chimie de la Matière Condensée de Paris for the NIR spectrophotometric measurements. The authors also thank Alba Nicolás-Boluda, Jean-Baptiste Bodin, Rachel Méallet-Renault, Gilles Clavier, and Nicolas Tsapis for their fruitful discussions.

**Conflicts of Interest:** The authors declare no conflict of interest. The funders had no role in the design of the study; in the collection, analyses, or interpretation of data; in the writing of the manuscript, or in the decision to publish the results.

## Appendix A

Appendix A demonstrates the suitability of the ultrasound detector frequency for the sample container.

The sample containers are tubes with an inner diameter of 0.2 mm. For an optically thin and infinitely long cylinder of water surrounded by water, the PA-generated ultrasound waves captured in the far-field are expected to have an acoustic spectrum, which has a magnitude proportional to [33]:

$$p(f) \propto \frac{J_1\left(2\pi \frac{a}{v_s^{water}} f\right)}{\sqrt{f}}, \quad \text{(A1)}$$

where $f$ is the ultrasound frequency, $J_1$ is the first order Bessel function, $a$ is the radius of the cylinder, and $v_s^{water}$ is the speed of sound in water. The first and highest peak of this acoustic spectrum is bounded by the first zero of Equation (A1):

$$f_{1st\ zero} \approx 3.83 \frac{v_s^{water}}{2\pi a}, \quad \text{(A2)}$$

For an optimal detector sensitivity, the first peak should cover the entire frequency bandwidth of the detector, and the first zero should slightly exceed the upper frequency of the ultrasound bandwidth of the detector. The upper frequency of our detector given by the clinical array manufacturer for pulse-echo ultrasound imaging is around 7 MHz. For $f_{1st\ zero} \geq 7$ MHz and $v_s^{water} = 1500$ m.s$^{-1}$, the maximal inner radius of the tube would be 130 μm. With $a$ = 100 μm, the nominal radius of our tubes, the theoretical spectrum is displayed in Figure 1b with the dashed blue curve. The solid black curve presents the ultrasound spectrum for the baseline-corrected signal of one tube. The entire bandwidth of the detector was indeed adequately excited by the chosen sample container.

**Appendix B**

Appendix B demonstrates the *in vitro* capability of the calibration method to operate in a scattering and absorbing environment such as biological tissues. This configuration could be used to verify the detectability of a contrast agent with the PAI system in a tissue-mimicking phantom, for instance.

A tissue phantom was implemented: the four PTFE tubes, placed in the same tube holder as described in Section 2.1.1., were surrounded with raw turkey breast (Figure A1a). The turkey breast was purchased from the local grocery store. Three 0.5 cm thick slices of tissue were placed on top of tubes 1 and 3 (Figures 1 and A1c), in between the tubes 1/3 and 2/4, and below tubes 2 and 4, respectively. Thereby, the tubes were embedded in a 1.5 cm thick tissue block, and the laser light was scattered and partially absorbed before reaching the tubes. The in vitro tissue phantom was immersed in water with 0.9% NaCl (kitchen salt) at 21.7 °C ± 0.2 °C.

Acquisitions were performed in the same condition as for the clear medium: the samples were injected in the tubes, and blank acquisitions were recorded with ultrapure water inside the tubes. The same data processing was applied to obtain the PA coefficient $\theta^{PA}(\lambda)$ of the sample (see Sections 2.1 and 2.2). The 100% and the 60% solutions of nigrosin and the 80% solution of $CuSO_4$ were measured. As for the clear medium (see Section 3.1.3.), the calibration solution was injected twice before and twice after the samples for a total of four acquisitions per tube, and the samples were injected twice.

Figure A1b,c presents the images without and with subtraction of the blank acquisition. The photoacoustic signal of the tissue was mainly removed by the subtraction process, enabling extraction of $A^{PA}$. Figure A1d displays the PA coefficient $\theta^{PA}(\lambda)$ of the three solutions and the SPP spectra. For the nigrosin samples, we verified that the PA and SPP spectra are superimposed (PGE equal to one). A PGE coefficient of 1.16 was found for the $CuSO_4$ solution, which matches with Section 3.2.2.

These results demonstrate that the calibration method can be used in scattering and absorbing media such as an *in vitro* tissue phantom.

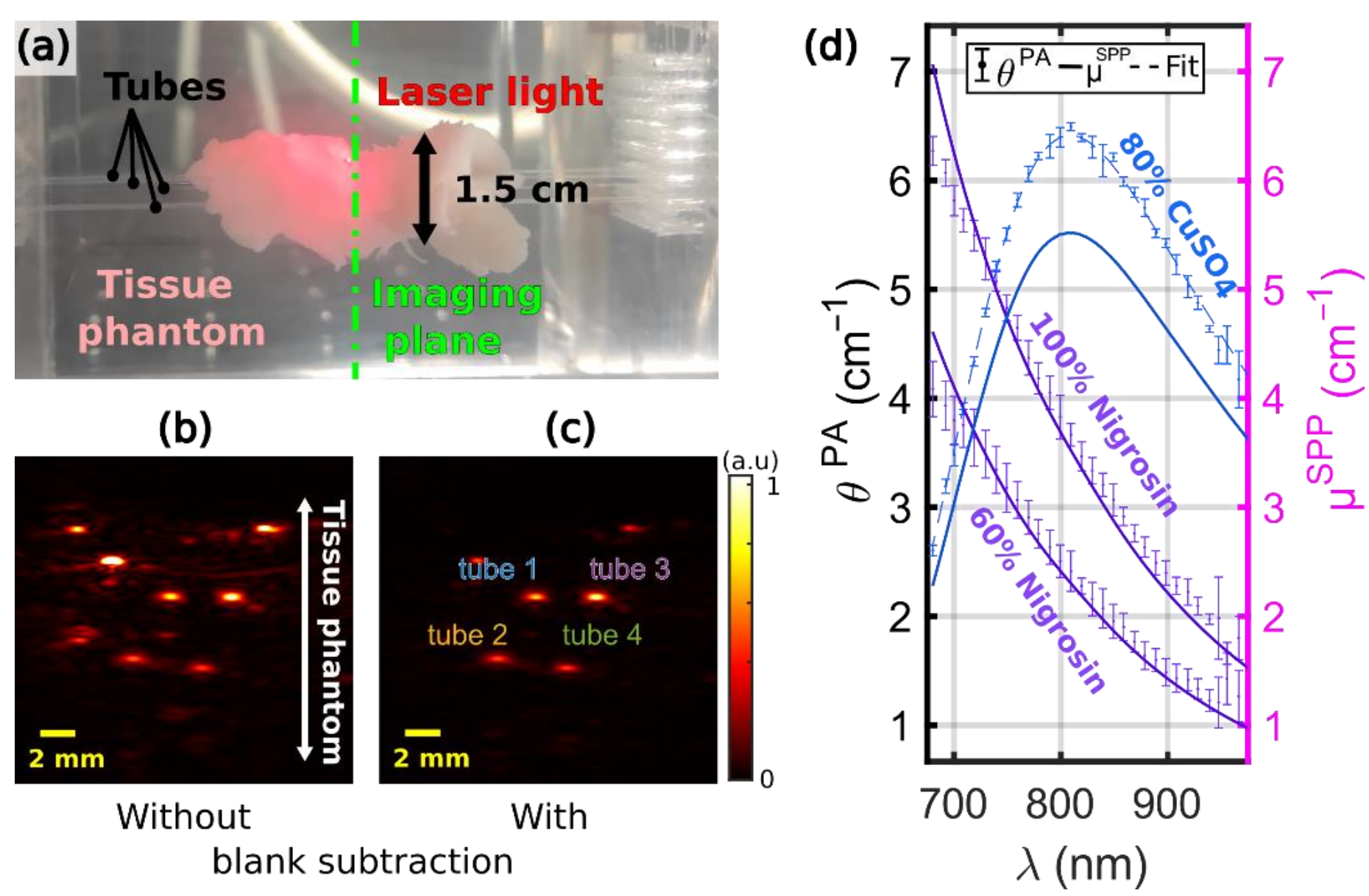

**Figure A1.** Performance of the calibration method with a tissue phantom. (**a**) Annotated picture of the tissue phantom. The tubes are surrounded by raw turkey breast. Three tissue layers (0.5 cm thick) form a tissue block with a total thickness of 1.5 cm: one layer above tubes 1/3, one layer between tubes 1/3 and tubes 2/4, and one layer below tubes 2/4. (**b**,**c**) PA images at 800 nm and with the tubes filled with the 100% solution of nigrosin. PA images (**b**) without and (**c**) with the subtraction of the blank acquisition (tubes filled with water) to the radio-frequency data before image reconstruction. (**b**,**c**) have the same color scale, which was set to correspond to the range of values in (**c**). (**d**) PA coefficients obtained for three solutions: the 100% and the 60% solutions of nigrosin (purple) and the 80% solution of $CuSO_4$ (blue). The solutions were measured twice in four tubes. The median ± MAD values of $\theta^{PA}(\lambda)$ were displayed with dot markers and error bars. The absorption coefficient $\mu^{SPP}(\lambda)$ is displayed with a solid line. The fit of $\mu^{SPP}$ to $\tilde{\theta}^{PA}$ is shown with a dashed line for the 80% solution of $CuSO_4$.

## References

1. Beard, P. Biomedical photoacoustic imaging. *Interface Focus* **2011**, *1*, 602–631. https://doi.org/10.1098/rsfs.2011.0028.
2. Taruttis, A.; Ntziachristos, V. Advances in real-time multispectral optoacoustic imaging and its applications. *Nat. Phot.* **2015**, *9*, 219–227. https://doi.org/10.1038/nphoton.2015.29.
3. Weber, J.; Beard, P.C.; Bohndiek, S.E. Contrast agents for molecular photoacoustic imaging. *Nat. Methods* **2016**, *13*, 639–650. https://doi.org/10.1038/nmeth.3929.
4. Liu, Y.; Bhattarai, P.; Dai, Z.; Chen, X. Photothermal therapy and photoacoustic imaging *via* nanotheranostics in fighting cancer. *Chem. Soc. Rev.* **2018**, *48*, 2053–2108. https://doi.org/10.1039/C8CS00618K.
5. Borg, R.E.; Rochford, J. Molecular Photoacoustic Contrast Agents: Design Principles & Applications. *Photochem. Photobiol.* **2018**, *94*, 1175–1209. https://doi.org/10.1111/php.12967.
6. Ho, C.J.H.; Balasundaram, G.; Driessen, W.; McLaren, R.; Wong, C.L.; Dinish, U.S.; Attia, A.B.E.; Ntziachristos, V.; Olivo, M. Multifunctional photosensitizer-based contrast agents for photoacoustic imaging. *Sci. Rep.* **2014**, *4*, 5342. https://doi.org/10.1038/srep05342.
7. Armanetti, P.; Flori, A.; Avigo, C.; Conti, L.; Valtancoli, B.; Petroni, D.; Doumett, S.; Cappiello, L.; Ravagli, C.; Baldi, G.; et al. Spectroscopic and photoacoustic characterization of encapsulated iron oxide super-paramagnetic nanoparticles as a new multiplatform contrast agent. *Spectrochim. Acta Part A Mol. Biomol. Spectrosc.* **2018**, *199*, 248–253. https://doi.org/10.1016/j.saa.2018.03.025.
8. Laufer, J.; Zhang, E.; Beard, P. Evaluation of absorbing chromophores used in tissue phantoms for quantitative photoacoustic spectroscopy and imaging. *IEEE J. Sel. Top. Quantum Electron.* **2010**, *16*, 600–607. https://doi.org/10.1109/JSTQE.2009.2032513.
9. Stahl, T.; Allen, T.; Beard, P. Characterization of the thermalisation efficiency and photostability of photoacoustic contrast agents. In *Photons Plus Ultrasound: Imaging and Sensing 2014*; SPIE: Bellingham, WA, USA, 2014; Volume 8943, p. 89435H. https://doi.org/10.1117/12.2039694.
10. Stahl, T. *Characterisation of Contrast Agents for Photoacoustic Imaging*; University College London: London, UK, 2016.
11. Fonseca, M.; An, L.; Beard, P.; Cox, B. Sulfates as chromophores for multiwavelength photoacoustic imaging phantoms. *J. Biomed. Opt.* **2017**, *22*, 1. https://doi.org/10.1117/1.JBO.22.12.125007.
12. Fuenzalida Werner, J.P.; Huang, Y.; Mishra, K.; Janowski, R.; Vetschera, P.; Heichler, C.; Chmyrov, A.; Neufert, C.; Niessing, D.; Ntziachristos, V.; et al. Challenging a Preconception: Optoacoustic Spectrum Differs from the Optical Absorption Spectrum of Proteins and Dyes for Molecular Imaging. *Anal. Chem.* **2020**, *92*, 10717–10724. https://doi.org/10.1021/acs.analchem.0c01902.
13. Pelivanov, I.; Petrova, E.; Yoon, S.J.; Qian, Z.; Guye, K.; O'Donnell, M. Molecular fingerprinting of nanoparticles in complex media with non-contact photoacoustics: Beyond the light scattering limit. *Sci. Rep.* **2018**, *8*, 14425. https://doi.org/10.1038/s41598-018-32580-2.

14. Schellenberg, M.W.; Hunt, H.K. Hand-held optoacoustic imaging: A review. *Photoacoustics* **2018**, *11*, 14–27. https://doi.org/10.1016/j.pacs.2018.07.001.
15. Kratkiewicz, K.; Manwar, R.; Zhou, Y.; Mozaffarzadeh, M.; Avanaki, K. Technical Considerations when using Verasonics Research Ultrasound Platform for Developing a Photoacoustic Imaging System. *Biomed. Opt. Express* **2021,** *12*, 1050-1084 . https://doi.org/10.1364/boe.415481.
16. Arconada-Alvarez, S.J.; Lemaster, J.E.; Wang, J.; Jokerst, J.V. The development and characterization of a novel yet simple 3D printed tool to facilitate phantom imaging of photoacoustic contrast agents. *Photoacoustics* **2017**, *5*, 17–24. https://doi.org/10.1016/j.pacs.2017.02.001.
17. Marczak, W. Water as a standard in the measurements of speed of sound in liquids. *J. Acoust. Soc. Am.* **1997**, *102*, 2776–2779. https://doi.org/10.1121/1.420332.
18. Xu, M.; Wang, L.V. Photoacoustic imaging in biomedicine. *Rev. Sci. Instrum.* **2006**, *77*, 041101. https://doi.org/10.1063/1.2195024.
19. Zhou, Y.; Yao, J.; Wang, L.V. Tutorial on photoacoustic tomography. *J. Biomed. Opt.* **2016**, *21*, 061007. https://doi.org/10.1117/1.jbo.21.6.061007.
20. Yao, D.-K.; Zhang, C.; Maslov, K.; Wang, L.V. Photoacoustic measurement of the Grüneisen parameter of tissue. *J. Biomed. Opt.* **2014**, *19*, 017007. https://doi.org/10.1117/1.JBO.19.1.017007.
21. Gehrung, M.; Bohndiek, S.E.; Brunker, J. Development of a blood oxygenation phantom for photoacoustic tomography combined with online $pO_2$ detection and flow spectrometry. *J. Biomed. Opt.* **2019**, *24*, 1. https://doi.org/10.1117/1.jbo.24.12.121908.
22. Petrova, E.; Ermilov, S.; Su, R.; Nadvoretskiy, V.; Conjusteau, A.; Oraevsky, A. Using optoacoustic imaging for measuring the temperature dependence of Grüneisen parameter in optically absorbing solutions. *Opt. Express* **2013**, *21*, 25077. https://doi.org/10.1364/oe.21.025077.
23. Kirchherr, A.K.; Briel, A.; Mäder, K. Stabilization of indocyanine green by encapsulation within micellar systems. *Mol. Pharm.* **2009**, *6*, 480–491. https://doi.org/10.1021/mp8001649.
24. Hale, G.M.; Querry, M.R. Optical Constants of Water in the 200-nm to 200-μm Wavelength Region. *Appl. Opt.* **1973**, *12*, 555. https://doi.org/10.1364/ao.12.000555.
25. Jain, P.K.; Lee, K.S.; El-Sayed, I.H.; El-Sayed, M.A. Calculated absorption and scattering properties of gold nanoparticles of different size, shape, and composition: Applications in biological imaging and biomedicine. *J. Phys. Chem. B* **2006**, *110*, 7238–7248. https://doi.org/10.1021/jp057170o.
26. Chen, Y.S.; Frey, W.; Aglyamov, S.; Emelianov, S. Environment-dependent generation of photoacoustic waves from plasmonic nanoparticles. *Small* **2012**, *8*, 47–52. https://doi.org/10.1002/smll.201101140.
27. Chen, Y.-S.; Zhao, Y.; Yoon, S.J.; Gambhir, S.S.; Emelianov, S. Miniature gold nanorods for photoacoustic molecular imaging in the second near-infrared optical window. *Nat. Nanotechnol.* **2019**, *14*, 465–472. https://doi.org/10.1038/s41565-019-0392-3.
28. Mantri, Y.; Jokerst, J.V. Engineering Plasmonic Nanoparticles for Enhanced Photoacoustic Imaging. *ACS Nano* **2020**, *14*, 9408–9422. https://doi.org/10.1021/acsnano.0c05215.
29. Schwartz, A.; Joline, L.; Snow, E.; Francke, S.; Marques, C.; Bernardes, R. Stabilizing Indocyanin Green for Use as Quantitative Standards. *Invest. Ophthalmol. Vis. Sci.* **2008**, *49*, 2116.
30. Wu, Y.; Zhang, H.K.; Kang, J.; Boctor, E.M. An economic photoacoustic imaging platform using automatic laser synchronization and inverse beamforming. *Ultrasonics* **2020**, *103*, 106098. https://doi.org/10.1016/j.ultras.2020.106098.
31. Rathod, V.T. A review of acoustic impedance matching techniques for piezoelectric sensors and transducers. *Sensors* **2020**, *20*, 4051. https://doi.org/10.3390/s20144051.
32. Mabuza, B.R.; Netshidavhini, N. The effect of a middle layer on ultrasonic wave propagating in a three-layered structure. In Proceedings of the Meetings on Acoustics ICA2013, Montreal, QC, Canada, 2–7 June 2013; Volume 19, p. 030076. https://doi.org/10.1121/1.4799683.
33. Khan, M.I.; Sun, T.; Diebold, G.J. Photoacoustic waves generated by absorption of laser radiation in optically thin cylinders. *J. Acoust. Soc. Am.* **1993**, *94*, 931–940. https://doi.org/10.1121/1.408195.